\pdfoutput=0
\documentclass[11pt]{article}
\usepackage{epsfig}
\usepackage{amsfonts}
\usepackage{amsmath}
\usepackage{bm,bbm}
\usepackage{cite}
\usepackage{cancel}
\usepackage{hyperref}
\allowdisplaybreaks[1]

\newcommand{\la}[1]{\label{#1}}
\newcommand{\be}{\begin{equation}}
\newcommand{\ee}{\end{equation}}
\newcommand{\ba}{\begin{eqnarray}}
\newcommand{\ea}{\end{eqnarray}}
\newcommand{\fig}{Fig.~}
\newcommand{\figs}{Figs.~}
\newcommand{\eq}{Eq.~}
\newcommand{\eqs}{Eqs.~}
\newcommand{\se}{Sec.~}
\newcommand{\ses}{Secs.~}
\newcommand{\nr}[1]{(\ref{#1})}
\newcommand{\nn}{\nonumber \\}

\renewcommand{\vec}[1]{{\bf #1}}

\newcommand{\tfr}[2]{{\textstyle \frac{#1}{#2}\,}}

\renewcommand{\eq}{eq.~}
\renewcommand{\eqs}{eqs.~}
\renewcommand{\se}{sec.~}
\renewcommand{\ses}{secs.~}
\renewcommand{\fig}{fig.~}
\renewcommand{\figs}{figs.~}
 
\newcommand{\Nc}{N_{\rm c}}

\newcommand{\rmO}{{\mathcal{O}}}

\def\lsi{\raise0.3ex\hbox{$<$\kern-0.75em\raise-1.1ex\hbox{$\sim$}}}
\def\gsi{\raise0.3ex\hbox{$>$\kern-0.75em\raise-1.1ex\hbox{$\sim$}}}

\newcommand{\sign}{\mathop{\mbox{sign}}}
\newcommand{\rmi}[1]{{\mbox{\scriptsize #1}}}
\newcommand{\rmii}[1]{{\mbox{\tiny\rm{#1}}}}
\newcommand{\rmiii}[1]{{\mbox{\tiny{$\scriptstyle{\rm#1}$}}}}

\newcommand{\Tint}[1]{{\hbox{$\sum$}\!\!\!\!\!\!\!\int\,}_{\!\!\!\!\raise-0.9ex\hbox{$\scriptstyle{#1}$}}}
\newcommand{\Tinti}[1]{{{\Sigma}\!\!\!\!\raise0.3ex\hbox{$\int$}_\rmii{${#1}$}}}
\newcommand{\bi}{\begin{itemize}}
\newcommand{\ei}{\end{itemize}}
\newcommand{\hide}[1]{ }

\newcommand{\deltabar}{\raise-0.02em\hbox{$\bar{}$}\hspace*{-0.8mm}{\delta}}
\newcommand{\ddeltabar}{\raise-0.18em\hbox{$\bar{}$}\hspace*{-0.8mm}{\delta}}
\renewcommand{\P}{\mathcal{P}}
\renewcommand{\H}{\mathcal{H}}

\newcommand{\X}{\mathcal{X}}
\newcommand{\Y}{\mathcal{Y}}

\newcommand{\T}{\rmii{$T$}}

\newcommand{\mpl}{m_\rmii{pl}} 

\newcommand{\hcoeff}{\mathcal{F}}
\newcommand{\E}{\mathcal{S}}
\newcommand{\der}{,}
\def\TAsc(#1,#2)(#3,#4,#5)%
{\SetWidth{2.0}\CArc(#1,#2)(#3,#4,#5)\SetWidth{1.0}}
\def\Lwidth{3}

\def\TAgl(#1,#2)(#3,#4,#5){\SetWidth{2.0}\PhotonArc(#1,#2)(#3,#4,#5){\Lwidth}%
{6.283 #3 mul 360 div #4 #5 sub #4 #5 sub mul sqrt mul Tdensity mul}%
\SetWidth{1.0}}
\def\TLgl(#1,#2)(#3,#4){\SetWidth{2.0}\Photon(#1,#2)(#3,#4){\Lwidth}
{#1 #3 sub #1 #3 sub mul #2 #4 sub #2 #4 sub mul add sqrt Tdensity mul}%
\SetWidth{1.0}}

\def\Lwidth{1.3}

\makeatletter \@addtoreset{equation}{section} \makeatother
\renewcommand{\theequation}{\arabic{section}.\arabic{equation}}
\makeatletter
\renewcommand\section{\@startsection {section}{1}{\z@}%
                                   {-5.5ex \@plus -1ex \@minus -.2ex}
                                   {2.3ex \@plus.2ex}%
                                   {\normalfont\large\bfseries}}
\renewcommand\subsection{\@startsection{subsection}{2}{\z@}%
                                     {-3.25ex\@plus -1ex \@minus -.2ex}%
                                     {1.5ex \@plus .2ex}%
                                     {\normalfont\normalsize\bfseries}}
\renewcommand\thesection {\@arabic\c@section}
\renewcommand\thesubsection   {\thesection.\@arabic\c@subsection}
\renewcommand{\@seccntformat}[1]{%
\csname the#1\endcsname.\hspace{1.0em}}
\makeatother

\begin{document}

\flushbottom

\begin{titlepage}

\begin{flushright}
October 2024
\end{flushright}
\begin{centering}
\vfill

{\Large{\bf
    Evolution of coupled scalar perturbations through \\[3mm]
    smooth reheating. I.~Dissipative regime
}} 

\vspace{0.8cm}

M.~Laine\hspace*{0.4mm}$^\rmi{a}_{ }$,
S.~Procacci\hspace*{0.4mm}$^\rmi{b}_{ }$,
A.~Rogelj\hspace*{0.4mm}$^\rmi{a}_{ }$

\vspace{0.8cm}

$^\rmi{a}_{ }${\em
AEC, 
Institute for Theoretical Physics, 
University of Bern, \\ 
Sidlerstrasse 5, CH-3012 Bern, Switzerland \\}

\vspace*{0.3cm}

$^\rmi{b}_{ }${\em
D\'epartement de Physique Th\'eorique, 
Universit\'e de Gen\`eve, \\ 
24 quai Ernest Ansermet, 
CH-1211 Gen\`eve 4, Switzerland \\}

\vspace*{0.8cm}

\mbox{\bf Abstract}
 
\end{centering}

\vspace*{0.3cm}
 
\noindent
If the inflaton is a heavy scalar field, it may equilibrate slower than 
some other degrees of freedom, e.g.\ non-Abelian gauge bosons. In this case, 
perturbations in the inflaton field and in a thermal 
plasma coexist from a given moment onwards. We derive a gauge-invariant set
of three coupled equations governing the time evolution of such a system. 
Despite singular coefficients, 
a reliable numerical solution can be obtained for a long 
time period, starting from phase oscillations inside the Hubble horizon,
and extending until acoustic oscillations in a radiation-dominated universe.   
Benchmarks are illustrated from a ``weak regime'', 
where perturbations have a quantum-mechanical origin but get dissipated 
by interactions with the plasma. Among applications of our formalism
could be inhomogeneity-induced nucleations in 
post-inflationary phase transitions, and 
the production of scalar-induced gravitational waves.

\vfill


\end{titlepage}

\tableofcontents

%
\section{Introduction and outline}
\la{se:intro}

While inflationary physics is traditionally discussed in the context of 
Cosmic Microwave Background (CMB) observations and large-scale structure
formation, recently also shorter length scales have come
under focus. One reason is the emergence of gravitational wave astronomy, 
opening up a huge frequency window, from pulsar time arrays
($f^{ }_0 \sim 10^{-8}_{ }$~Hz) up to tabletop experiments 
($f^{ }_0 \sim 10^{12}_{ }$~Hz). Another is the fashionable idea
that primordial black holes, generated from the collapse of 
distinctive short-wavelength features in the density
power spectrum, could constitute part of dark matter.  

The physics of short length scales is less well understood
than that of longer ones. One reason is that such modes exit the 
Hubble horizon later than the CMB modes, and re-enter it earlier.
Therefore they are more affected by 
the complicated physics taking place when
inflation ends and a thermal plasma is generated~\cite{reheat,preheat}. 
Furthermore, there is plenty of time for sub-horizon causal
phenomena to take place, whereby density and metric 
perturbations are not frozen, but rather couple
to the hydrodynamic modes of a plasma. 

A concrete example of the challenges met in treating 
density perturbations in the presence of 
hydrodynamic phenomena, is offered by the framework of warm inflation. 
It posits that a thermal plasma could have existed already
during inflation
(see, e.g.,\ refs.~\cite{warm0,warm1,warm2} for reviews, 
and refs.~\cite{R_1,R_2,rr,R_3,mehrdad,rosa,ballesteros,freese}
for methods).
The degrees of freedom are the 
fluctuations of a scalar field, $\delta\varphi$; 
of a temperature, $\delta T$; 
of a plasma flow velocity, $v$; 
and of the spacetime metric, $\delta g^{ }_{\mu\nu}$.

Studying warm inflation literature, or inflation literature
more generally, at least two technical challenges stand out. 
One is gauge fixing. While in principle 
there is nothing wrong with gauge fixing, the problem
is that many different gauges are being made use of, complicating
comparisons. Sometimes different gauge choices are employed even 
within a single computation, leading easily to confusion. 

The problems with gauge fixing can be illustrated by a recipe that 
is often used as a rough estimate (cf.,\ e.g.,\ ref.~\cite{R_4}). 
Suppose that $\delta\varphi$ is 
first solved as if there were no $\delta g^{ }_{\mu\nu}$ present. 
{}From the corresponding power spectrum, 
$ \P^{ }_{\delta\varphi} $, a gauge-invariant curvature
power spectrum, 
$
 \P^{ }_{\mathcal{R}_\varphi}
$, 
is estimated as 
$
 \P^{ }_{\mathcal{R}_\varphi}
    \simeq
 (H / \dot{\bar{\varphi}})^2_{ }
 \,\P^{ }_{\delta\varphi}
$, 
where $\mathcal{R}^{ }_\varphi$ is from \eq\nr{def_R_varphi}, 
$H$ is the Hubble rate at horizon exit,
and $\bar\varphi$ is the inflaton background.
The problem is that removing
$\delta g^{ }_{\mu\nu}$ from the equation of motion of $\delta\varphi$,
and from the relation between $\delta\varphi$ 
and $\mathcal{R}^{ }_\varphi$, 
require {\em different} gauge choices. In fact, even in vacuum, 
the equation of motion satisfied by $\delta\varphi$ in the 
absence of $\delta g^{ }_{\mu\nu}$  
(cf.\ \eq\nr{langevin_vac_3}),
is different from that satisfied by $\mathcal{R}^{ }_\varphi$
(cf.\ \eq\nr{eq_hatQ_varphi}), 
as soon as we 
depart from pure de Sitter expansion
(cf.\ \eq\nr{extras}). 
Concretely, this suggests that
if CMB observables are computed
as an expansion in slow-roll parameters, 
subleading terms differ between the recipe
and a gauge-invariant computation.

The second challenge is that, due to the traditional focus on
CMB observables, many studies restrict themselves 
to the slow-roll stage of inflation, 
and to the computation of $\mathcal{R}^{ }_\varphi$ during that period. 
Then it is asserted that after inflation, 
$\varphi$ disappears, and $\mathcal{R}^{ }_\varphi$ should be replaced
by another curvature perturbation. But there are many of them
(cf.,\ e.g.,\ \eqs\nr{def_R_v} and \nr{def_R_T}), so which one? 
Furthermore, $\mathcal{R}^{ }_\varphi$ contains the ratio 
$\delta\varphi / \dot{\bar\varphi}$, and even if the amplitudes of 
both $\delta\varphi$ and $\dot{\bar\varphi}$ decrease, it is
not obvious how the ratio behaves. 

The purpose of the present paper is to clarify
the simultaneous treatment of 
$\delta\varphi$, $\delta T$, $v$, and $\delta g^{ }_{\mu\nu}$. 
Specifically, we work in a general gauge~\cite{bardeen,mfb,hn}, 
and establish 
a closed set of equations directly for $\mathcal{R}^{ }_\varphi$
and two other gauge-invariant perturbations
(cf.\ \eqs\nr{dot_R_varphi}--\nr{dot_S_T}), which do not rely
on the slow-roll approximation. We study how 
$\mathcal{R}^{ }_\varphi$ behaves when $\bar\varphi$ undergoes rapid
oscillations, and when the energy density associated with 
$\dot{\bar\varphi}^2_{ }$ decreases below that of radiation. 
We solve the equations all
the way to a radiation-dominated era, when 
modes re-enter the Hubble horizon and start to undergo
acoustic oscillations.

We remark that our ``radiation'' plasma is assumed to be 
composed of degrees of freedom {\em separate} from the inflaton field. 
For instance, it could consist of non-Abelian gauge bosons, which
are known to equilibrate fast~\cite{equil}. However the 
gauge bosons' precise identity is unimportant --- they could be those of a 
``dark sector'', or directly of the Standard Model.

Given that the topic turned out to require an extensive discussion, 
we have decided to split our considerations into parts. In this 
first part, we study the dissipative regime, where an important
role is played by a damping coefficient $\Upsilon$, removing
energy from the inflaton and transferring it to the plasma. 
Roughly speaking, this coincides with the so-called
weak regime of warm inflation, with $\Upsilon \ll H$. 
Then the inflaton field itself does {\em not} equilibrate.  
In a future second part, which requires a solution of stochastic differential 
equations, the effects of thermal fluctuations are included. 
The fluctuation regime coincides with the so-called strong regime
of warm inflation, with $\Upsilon \gg H$. 
Then the inflaton field may thermalize. 
Our discussion is meant to be model-independent, even if for
numerical illustrations benchmarks are chosen. 
Specific model studies, 
either for warm inflation, or for short-distance
features originating close to the heating-up period, are relegated
to subsequent works. 

\vspace*{3mm}

Our presentation is organized as follows. 
In~\se\ref{se:defs}, we define first-order metric perturbations, 
the energy-momentum tensor of the full system, and the dynamics
governing the inflaton field.  
Inserting these ingredients into the Einstein equations, 
a set of coupled equations for three curvature perturbations  
is obtained in~\se\ref{se:curvature}. 
In \se\ref{se:subtleties}, we 
show how initial conditions can be fixed at early times; 
singularities can be avoided at intermediate times; and
simplified equations can be obtained at late times.
Numerical illustrations can be found in~\se\ref{se:numerics},
before we conclude with an outlook in~\se\ref{se:concl}.
In appendix~\ref{se:quantum}, the canonical quantization of 
a scalar field in an expanding background is briefly reviewed.

%
\section{Basic definitions}
\la{se:defs}

%
\subsection{Metric perturbations}

In analyses of inflationary fluctuations, 
it is conventional to make
use of gauge invariance, and select a specific gauge, 
in order to simplify the equations. 
Here, we choose another philosophy, 
carrying out the computations in a general gauge. 
One advantage is that gauge redundancy can then be exploited 
as a powerful crosscheck of the computations. 

Consider metric perturbations which behave as 
scalar quantities under transverse rotations. 
There are four of them, denoted in the following by 
$h^{ }_0$, $h$, $h^{ }_\rmii{D}$, and $\vartheta$.\footnote{%
 Our notation follows ref.~\cite{sp}, 
 where the decomposition 
 of cosmological perturbations into
 scalar, vector and tensor parts is reviewed, together
 with their behaviour under coordinate transformations
 (see also ref.~\cite{hks}). 
 } 
In terms of these, 
the metric tensor and its inverse can be expressed 
as\hspace*{0.3mm}\footnote{%
 This corresponds to the ($-$$+$$+$$+$) signature. 
 } 
\ba
 g^{ }_{\mu\nu} & \equiv &  
 a^2_{ }
 \left(
  \begin{array}{cc} 
   -1 - 2 h^{ }_0 & \partial^{ }_i h \\ 
   \partial^{ }_i h & 
   ( 1 - 2 h^{ }_\rmii{D}) \delta^{ }_{ij} + 
   2 \bigl( 
     \partial^{ }_i \partial^{ }_j - \delta^{ }_{ij} \frac{\nabla^2}{3}
     \bigr) \vartheta
  \end{array} 
 \right) 
 \;, \la{pert_metric} \\[2mm]
 g^{\mu\nu}_{ } & = &  
 a^{-2}_{ }
 \left(
  \begin{array}{cc} 
   -1 + 2 h^{ }_0 & \partial^{ }_i h \\ 
   \partial^{ }_i h & 
   ( 1 + 2 h^{ }_\rmii{D}) \delta^{ }_{ij} - 
   2 \bigl( 
     \partial^{ }_i \partial^{ }_j - \delta^{ }_{ij} \frac{\nabla^2}{3}
     \bigr) \vartheta
  \end{array} 
 \right) + \rmO(\delta^2) 
 \;, \la{pert_inv_metric}
\ea
where $\delta$ denotes generic first-order perturbations. 
The energy-momentum tensor is also perturbed (cf.\ \se\ref{sss:Tmunu}), 
and subsequently Einstein equations are imposed, 
$
 {G}^{\mu}_{ }{}^{ }_{\nu}
 = 8 \pi G 
 {T}^{\mu}_{ }{}^{ }_{\nu}
$, 
where $G \equiv 1/\mpl^2$, with 
$\mpl^{ } = 1.22091 \times 10^{19}_{ }$~GeV.

To motivate a nomenclature,
let us inspect what happens if we were to 
perturb the overall scale factor $a$. 
We focus on the spatial part, given that the conformal time, $\tau$, 
may be replaced by the physical time, $t$, in which case no~$a$ 
appears in~$g^{ }_{00}$.
Then, from the spatial part 
$g^{ }_{ij} = a^2_{ }\delta^{ }_{ij} + \rmO(\delta)$ of \eq\nr{pert_metric},  
we get  
$
 \delta g^{ }_{ij} \supset  (2 a\, \delta a)\, \delta^{ }_{ij}
$. 
Comparing with \eq\nr{pert_metric}, we see that
\be
 \frac{\delta a}{a} \bigg|^{ }_\rmi{spatial} 
 \; \Leftrightarrow \; 
 - \biggl( h^{ }_\rmii{D} + \frac{\nabla^2\vartheta}{3} \biggr)
 + \rmO(\delta^2)  
 \;. 
\ee
This particular combination turns out {\em not} to be gauge invariant, 
i.e.\ invariant under coordinate transformations. However, if we add
a matter part, gauge-invariant combinations can be defined, and
in the following we call them {\em curvature perturbations}, 
\be
 \mathcal{R}^{ }_x \; \equiv \; 
 - \biggl( h^{ }_\rmii{D} + \frac{\nabla^2\vartheta}{3} \biggr)
 + \mbox{(contribution of matter of type $x$)}
 \;. \la{def_R_x}
\ee

%
\subsection{Energy-momentum tensor}
\la{sss:Tmunu}

For the right-hand side of the Einstein equations, we need to specify 
the energy-momentum tensor of the system. 
We assume that it takes the form 
\be
 T^{\mu\nu}_{  }
   \; \equiv \; 
 T^{\mu\nu}_\rmi{ideal}
   \; \equiv \; 
 \varphi^{,\mu}_{ }\varphi^{,\nu}_{ }
 - \frac{ g^{\mu\nu}_{ }  \varphi^{ }_{,\alpha} \varphi^{,\alpha}_{ } }{2} 
 + ( e + p ) \,  u^{\mu}_{ }u^{\nu}_{ }+ p \, g^{\mu\nu}_{ }
 \;, \la{Tmunu_mixed}
\ee
where $\varphi$ denotes our inflaton field, 
$(...)^{ }_{,\alpha}$ 
a usual derivative, 
and (later on) 
$(...)^{ }_{;\alpha}$
a covariant derivative. 
We have included only the ideal
part of the plasma contribution to the energy-momentum tensor, 
i.e.\ no viscosities; viscous effects are however 
generated dynamically by the equations we solve, as explained 
at the beginning of \se\ref{ss:inflaton}.  
We assume the plasma to consist of only one fluid, with
an equilibration rate $\Gamma \gg \Upsilon$, where 
$\Upsilon$ represents the equilibration rate of 
$\varphi$, to be introduced in \eq\nr{varphi_eq}. 

The energy density and pressure
appearing in \eq\nr{Tmunu_mixed} include the contributions of plasma
radiation ($e^{ }_r$, $p^{ }_r$) and of the inflaton effective
potential ($V$), which is itself a function of~$\varphi$ 
and the temperature, $T$. 
We stress that our $e$ and $p$ only include non-derivative terms, 
whereas the derivatives acting on $\varphi$ are kept separate in 
\eq\nr{Tmunu_mixed}. 
This is because the derivatives have a different tensor structure
from the radiation part: $\varphi^{,\mu}_{ }$ does not need to  
align with the fluid velocity $u^\mu_{ }$. This becomes important
when we go to perturbations, so that spatial derivatives also 
play a role. 

The inflaton field and temperature are expanded around 
homogeneous background values, 
$\bar\varphi$ and $\bar T$, as 
\be
 \varphi \;\equiv\; \bar\varphi + \delta\varphi
 \;, \quad
 T \;\equiv\; \bar T + \delta T
 \;. \la{inflaton}
\ee
Likewise, the energy density and pressure are expressed as
$
 e = \bar{e} + \delta e
$
and
$
 p = \bar{p} + \delta p
$,
respectively.
To simplify the notation, we subsequently substitute 
$\bar{p}^{ }_r \to p^{ }_r$, 
$\bar{e}^{ }_r \to e^{ }_r$,
$\bar{V} \to V$, and  
$\bar T \to T$,  
whenever there is no danger of confusion.

As the effective potential 
corresponds to free energy density (minus the pressure), the 
background pressure reads
\be
 \bar{p} \; \equiv \; p^{ }_r - V
 \;, \quad
 p^{ }_r \; \equiv \; p^{ }_r(\bar T)
 \;, \quad
 V \; \equiv \; V(\bar\varphi,\bar T)
 \;. 
 \la{p_mixed}
\ee
If we take a time derivative of the average pressure, 
$(...)'_{ } \equiv \partial^{ }_\tau (...)$, 
or consider pressure perturbations, then there are two sources, 
\be
 \bar p\hspace*{0.2mm}' =
   \bar{s}\, T'
 - V^{ }_{\der\varphi}\, \bar \varphi\hspace*{0.4mm}'
 \;, \quad
 \delta p = 
   \bar{s}\, \delta T
 - V^{ }_{\der\varphi}\, \delta\varphi
 \;, \la{p_perturbations}
\ee
where 
parametric derivatives are abbreviated as  
$(...)^{ }_{\der x} \equiv 
\partial^{ }_x (...)$,
and the entropy density has been defined as
$
  \bar{s} \; \equiv \; \bar p^{ }_{\der\T}
 \; = \; p^{ }_{r\der\T} - V^{ }_{\der\T} 
$. 

The energy density is 
a Legendre transform of the pressure, 
\be
 \bar{e} \; = \; T \bar{p}^{ }_{\der\T} - \bar{p} 
 \; = \; e^{ }_r  + V - T V^{ }_{\der\T}
 \;, \quad
 e^{ }_r \; \equiv \; T p^{ }_{r\der\T} - p^{ }_r 
 \;. \la{e_mixed}
\ee
Its partial temperature derivative is related to the entropy 
density, and is also known as the heat capacity, 
$
 \bar{c} 
 \,\equiv\, 
 \bar{e}^{ }_{\der\T} 
  = 
 T \bar{s}^{ }_{\der\T} 
  =
  c^{ }_r - T V^{ }_{\der\T\T} 
$, 
$
 c^{ }_r 
 \,\equiv\, 
 e^{ }_{r\der\T}\,
$. 
The time derivative and the perturbation 
of the energy density take the form\hspace*{0.4mm}\footnote{%
 For generality, we keep the terms $V^{ }_{\der\varphi\T}$
 in the equations for the moment, however they tend to be 
 dangerous from the point of view of inflationary 
 phenomenology~\cite{warm0}, and indeed they will be absent
 from our numerical benchmarks in \se\ref{se:numerics}, in which the 
 potential is assumed temperature-independent. \la{fn:warm0}  
 }
\be
 \bar e\hspace*{0.2mm}' =
   \bar{c}\, T'
 + ( V^{ }_{\der\varphi}
 - T V^{ }_{\der\varphi\T})\, \bar \varphi\hspace*{0.4mm}'
 \;, \quad
 \delta e = 
   \bar{c}\, \delta T
 + ( V^{ }_{\der\varphi} - T V^{ }_{\der\varphi\T})\, \delta\varphi
 \;, \la{e_perturbations}
\ee
respectively. 
We also note that 
the combination appearing in \eq\nr{Tmunu_mixed} is 
proportional to the entropy density, 
\be
 \bar{e} + \bar{p} \; = \; T \bar{s}
 \; = \; e^{ }_r + p^{ }_r - T V^{ }_{\der\T} 
 \;. \la{e_plus_p}
\ee
In other words, with our definition 
$e+p$ plays a role only for thermal systems.

Inserting the metric tensor from
\eqs\nr{pert_metric} and \nr{pert_inv_metric} into  
\eq\nr{Tmunu_mixed}, and expanding the velocity 
around a homogeneous and isotropic background value,
$
 \bar u^{\mu}_{ } \equiv a^{-1}_{ }(1,\vec{0})
$,
$
 \bar{u}^{ }_\mu = a (-1,\vec{0})
$, 
non-diagonal first-order perturbations originate 
from $\delta u^i_{ }= a^{-1}_{ } v^i_{ }$.
The scalar velocity perturbation is parametrized by a velocity potential as
\be
 v^i_{ }\; \equiv \; v^{ }_i \; \equiv \; - v^{ }_{,i}
 \;,
\ee
so that 
\be
 u^\mu_{ } = a^{-1}_{ }\bigl( 1-h^{ }_0,-v^{ }_{,i} \bigr) + \rmO(\delta^2)
 \;, \quad
 u^{ }_\mu = a \bigl( -1-h^{ }_0,(h - v)^{ }_{,i} \bigr)  + \rmO(\delta^2)
 \;. \la{u_mu} 
\ee
Subsequently we write the energy-momentum tensor as 
\be
 {T^{\mu}_{ }}^{ }_\nu = 
 {\bar T^{\mu}_{ }}{}^{ }_\nu + 
 \delta {T^{\mu}_{ }}^{ }_\nu + 
 \rmO(\delta^2)
 \;, 
\ee
where the non-vanishing background values are 
\be
 {\bar T^{0}_{ }}{}^{ }_0  
 = 
 - \, 
 \biggl[
  \frac{(\bar\varphi ')^2_{ }}{2 a^2} + \bar e 
 \biggr] 
 \;, \quad
 {\bar T^{i}_{ }}{}^{ }_j
 = 
 \delta^{ }_{ij}
 \,  
 \biggl[
  \frac{(\bar\varphi ')^2_{ }}{2 a^2} + \bar p 
 \biggr] 
 \;. 
\ee
For the perturbations, the inflaton kinetic terms give
\be
  \delta {T^{\mu}_{ }}^{ }_\nu \, |^{ }_{\varphi} 
 = 
 \frac{ \bar\varphi' }{a^2_{ }} \, 
 \left(\, 
   \begin{array}{cc}
     h^{ }_0\, \bar\varphi' - \delta \varphi'
     & 
    -\delta\varphi^{ }_{,j} \\  
     h^{ }_{,i}\, \bar\varphi'  + \delta\varphi^{ }_{,i}    &
     \quad \delta^{ }_{ij}\,
     \bigl[ 
     - h^{ }_0\, \bar\varphi' + \delta \varphi'
     \bigr]
   \end{array}
 \,\right)
 \;, \la{delta_Tmunu_varphi}
\ee
whereas the radiation part yields
\be
  \delta {T^{\mu}_{ }}^{ }_\nu \, |^{ }_{r} 
 = 
 \left(\, 
   \begin{array}{cc}
    -\delta e & 
   (\bar{e} + \bar{p})\, (h-v)^{ }_{,j} \\ 
   (\bar{e} + \bar{p})\, v^{ }_{,i} & 
   \delta^{ }_{ij} \, \delta p 
   \end{array}
 \,\right)
 \;, \la{delta_Tmunu_r}
\ee
with $\delta p$ and $\delta e$ given 
in \eqs\nr{p_perturbations} and \nr{e_perturbations}, respectively. 
The background Einstein equations read
$
 {\bar G}^{\mu}_{ }{}^{ }_{\nu}
 = 8 \pi G 
 {\bar T}^{\mu}_{ }{}^{ }_{\nu}
$,
and the first-order perturbations satisfy
$
 \delta {G}^{\mu}_{ }{}^{ }_{\nu}
 = 8 \pi G 
 \delta {T}^{\mu}_{ }{}^{ }_{\nu}
$. 

We end this section by noting that in a practical computation, 
some of the information present in the Einstein equations can be 
more usefully represented through the energy-momentum conservation
equations, 
$
 {T^\mu_{ }}^{ }_{\nu;\mu} = 0
$.
In particular, this yields the radiation evolution equations, 
as will be discussed in \se\ref{sss:combined_R} 
(cf.\ \eqs\nr{dTmunu_i} and \nr{dTmunu_0}). 
It is important to stress 
that even though the inflaton field and the radiation component
separately feel dissipation, as specified by \eq\nr{varphi_eq} below, 
the overall energy-momentum is conserved. Therefore, the 
dissipative coefficient is absent from the overall energy-momentum
conservation equations (see, e.g., \eq\nr{bg_Tmunu}).
Once we subtract the dissipative inflaton equation from total
energy-momentum conservation, the radiation evolution equation 
also becomes dissipative (see, e.g., \eq\nr{eom_plasma}).

%
\subsection{Dynamics of the inflaton field}
\la{ss:inflaton}

The Einstein equations do not contain enough information to 
fix the time evolutions of all the quantities that we have introduced.
To close the system, 
one further equation is needed, which we introduce in this section. 

The dynamics of radiation and $\varphi$ are coupled through 
a Langevin equation, parametrized by a friction coefficient $\Upsilon$. 
This coupling induces dissipative effects to the system, 
notably a shear viscosity~\cite{hydro,scan}
(this happens at 2nd order in perturbations). 
The result for the shear viscosity is inversely 
proportional to~$\Upsilon$;
if $\Upsilon$ is small
and the inflaton fluctuations are not exponentially suppressed, 
then this dynamical contribution may be
larger than the effects
from the self-interactions of the plasma.
In the following we work under the assumption that the largest 
contributions to viscosities originate 
dynamically from~$\Upsilon$. 
Then we can employ the ideal form in 
\eq\nr{Tmunu_mixed} for the ``bare'' overall energy-momentum tensor. 

In a general background, the Langevin equation for a minimally
coupled\hspace*{0.4mm}\footnote{%
 We assume a minimal coupling for simplicity, and 
 because non-minimal couplings 
 should not appear if the inflaton
 field is a periodic (angular) variable, 
 like in ``natural'' inflation~\cite{ai}. 
 } 
scalar field takes the form 
\be
 {\varphi^{;\mu}_{ }}^{ }_{;\mu} - \Upsilon \, u^{\mu}_{ } \varphi^{ }_{,\mu}
  - V^{ }_{\der\varphi}  
  + \varrho 
 \; = \; 0
 \;. \la{varphi_eq} 
\ee
The coefficient $\Upsilon$ is responsible
for dissipating energy from $\varphi$ to the plasma;
the potential~$V$ describes 
the mass and self-interactions of $\varphi$; 
the noise $\varrho$ accounts 
for random fluctuations that transport energy from
the plasma to~$\varphi$. 

In general,  
$\Upsilon$ depends in a non-polynomial way 
on the plasma temperature~$T$, and
includes also a $T$-independent vacuum term.
Its $\varphi$-independent part can be expressed  
in terms of 
a 2-point correlation function of medium properties~\cite{warm}, 
but we do not need this information here. 
More generally, $\Upsilon$ also depends on $\varphi$~\cite{db}, 
so we allow for such a dependence. 

The autocorrelator of the noise is normally assumed to take the form 
\be
 \bigl\langle\,
    \varrho(\X) \varrho(\Y)
 \,\bigr\rangle \; = \; 
 \frac{\Omega \, \delta^{(4)}_{ }(\mathcal{X-Y})}{\sqrt{-\det g^{ }_{\mu\nu}}} 
 \;, \quad
 \X \; \equiv \; (x^0_{ },\vec{x})
 \;, \la{varphi_noise}
\ee
where the division in the numerator guarantees that the result
is a scalar quantity after integrating over 
spacetime ($\int^{ }_\X \sqrt{-\det g^{ }_{\mu\nu}} $).
However, in the present paper we seldom refer to \eq\nr{varphi_noise}, 
and make no assumptions about the coefficient $\Omega$.  
We rather keep $\varrho$ as a general source term 
for the evolution of $\varphi$, assuming only that it is 
a small quantity, of $\rmO(\delta\varphi)$. 

A few comments on the range of validity
of \eq\nr{varphi_eq} are in order. 
One is that even if we treat 
\eq\nr{varphi_eq} as a classical equation, 
quantum mechanics can be restored at linear order in perturbations
(cf.\ appendix~\ref{se:quantum}).
Second, the vacuum Klein-Gordon equation is obtained  
by omitting $\Upsilon$ and $\varrho$. Third, 
\eq\nr{varphi_eq} does {\em not} 
assume~$\varphi$ to be thermalized. A fluctuation-dissipation 
relation can be established  
between $\Upsilon$ and $\Omega$, from the assumption that $\varphi$
should thermalize at asymptotically large times,  
however we make no such assumption here. 

Inserting the metric and velocity perturbations from 
\eqs\nr{pert_metric}, \nr{pert_inv_metric} and \nr{u_mu}, 
the kinetic terms in \eq\nr{varphi_eq} become
\ba
 {\varphi^{;\mu}_{ }}^{ }_{;\mu} 
 & = & 
 \frac{1 - 2 h^{ }_0}{a^2_{ }}
 \,\Bigl[\,
  -\varphi'' + \nabla^2_{ }\varphi
 + \bigl( h_0' + 3 h_\rmii{D}' + \nabla^2_{ } h - 2\H \bigr) \varphi' 
 + \rmO(\delta^2_{ }\hspace*{0.2mm}) 
 \,\Bigr]
 \;, \la{varphi_kinetic} \\ 
 u^{\mu}_{ } \varphi^{ }_{,\mu} & = &
 \frac{ 1 - h^{ }_0}{a}
 \,\bigl[\, 
   \varphi' + \rmO(\delta^2_{ }\hspace*{0.2mm})
 \,\bigr]
 \;, \la{varphi_friction}
\ea
where
$
 \H \equiv a'/a
$
is the Hubble rate in conformal time.  
The terms not shown are of second order in small quantities, 
and not needed in the present study. 

%
\section{Curvature and isocurvature perturbations}
\la{se:curvature}

The first-order perturbations introduced in \se\ref{se:defs} are 
gauge dependent, i.e.\ they change in coordinate transformations.
Here we show how gauge-invariant linear combinations can be constructed, 
and derive the equations that govern their evolution.

%
\subsection{General considerations}
\la{sss:general}

Our basic variables are the metric $g^{ }_{\mu\nu}$, 
the inflaton field $\varphi$, 
and the plasma temperature and scalar velocity 
potential, $T$ and $v$. Their 
average values are parametrized by $\bar\Phi \equiv \{a, \bar\varphi, T\}$, 
and their fluctuations by 
$
 \delta \Phi \equiv \{ h^{ }_0,h,h^{ }_\rmii{D},\vartheta,\delta\varphi,
 \delta T, v \}
$.
At first order, there are 7 variables, but also 7 equations, namely 
the field equation for $\delta \varphi$, 
the components 
${{}^0_{ }}^{ }_0$, 
${{}^0_{ }}^{ }_i$,
${{}^i_{ }}^{ }_j |^{ }_\rmi{traceless part}$, 
${{}^i_{ }}^{ }_j |^{ }_\rmi{trace part}$
of the Einstein equations
$\delta {G^\mu_{ }}^{ }_\nu =  8 \pi G \delta {T^\mu_{ }}^{ }_\nu$, 
and the components 
$\nu =0$ and $\nu = i$ of 
the energy-momentum conservation equations
$
 \delta {T^\mu_{ }}^{ }_{\nu;\mu} = 0
$.
All of these are covariant under gauge transformations, 
\ba
 \tilde h^{ }_0 & = & h^{ }_0 - {\xi^0_{ }}' - \H \xi^0_{ }
 \;, \la{gauge1} \\[2.5mm]
 \tilde{h} & = & h + \xi^0_{ } + \xi'
 \;, \\[0.5mm] 
 \tilde h^{ }_\rmii{D} & = & h^{ }_\rmii{D} + 
 \H \xi^0_{ }  - \frac{\nabla^2\xi}{3}
 \;,   \\[2mm]
 \tilde\vartheta & = & \vartheta + \xi 
 \;, \\[2mm] 
 \delta\tilde\varphi & = & \delta\varphi - \bar{\varphi}' \xi^0_{ } 
 \;, \\[2mm] 
 \delta\tilde T & = & \delta T - T' \xi^0_{ } 
 \;, \\[2mm]
 \tilde v & = & v + \xi'
 \;,  \la{gauge7}
\ea
where $\xi^0_{ }$ and $\xi$ are arbitrary 
functions parametrizing temporal and spatial coordinate
shifts, respectively. 
Given that there are 2 gauge parameters, the 7 perturbations can be
re-grouped into 5 gauge-invariant subsets. 
Equivalently, there are only 5 independent equations. 

Now, linearizing a full equation of the form 
$ 
 \mathcal{D}(\Phi) = \varrho
$
after inserting
$
 \Phi = \bar{\Phi} + \delta \Phi + \rmO(\delta^2)
$, 
we are faced with 
\be
 \mathcal{D}(\bar\Phi) = 0 
 \;, \quad
 \mathcal{D}'(\bar\Phi)\, \delta\Phi = \varrho
 \;.  \la{general}
\ee
Perturbations can formally be solved for\hspace*{0.4mm}\footnote{%
 To be precise, this discussion assumes that quantum fluctuations
 have been represented through the formalism of stochastic inflation
 and are thus contained in $\varrho$ (cf.,\ e.g.,\ ref.~\cite{rr}).
 We do not otherwise make use of the stochastic formalism, however. 
 } 
as 
$
 \delta\Phi = [\mathcal{D}'(\bar\Phi)]^{-1}_{ } \varrho
$.
The general power spectrum is a correlation matrix, 
$
 \mathcal{P}^{ }_{\delta\Phi^{}_i \delta\Phi^{ }_j}
 \; = \; 
 \{ 
   [\mathcal{D}'(\bar\Phi)]^{-1}_{ } 
     \langle \varrho \varrho^T_{ } \rangle 
   [\mathcal{D}'(\bar\Phi)]^{-1T}_{ } 
 \}^{ }_{ij}
$.
A curvature power spectrum is obtained by projecting this 
according to the choice made in \eq\nr{def_R_x}, 
\be
 \mathcal{P}^{ }_{\mathcal{R}^{ }_x}
 \; = \; 
 e^{T}_{\mathcal{R}^{ }_x}
   [\mathcal{D}'(\bar\Phi)]^{-1}_{ } 
     \langle \varrho \varrho^T_{ } \rangle 
   [\mathcal{D}'(\bar\Phi)]^{-1T}_{ } 
 e^{ }_{\mathcal{R}^{ }_x}
 \;, \la{general_P_Rx}
\ee
where $ e^{ }_{\mathcal{R}^{ }_x} $ defines the weights with 
which various perturbations $\delta\Phi^{ }_i$ appear 
in $ \mathcal{R}^{ }_x $. 

In concrete terms, we define the curvature perturbations as 
\ba
 \mathcal{R}^{ }_\varphi
 & \equiv & 
 - \biggl( 
  h^{ }_\rmii{D} + \frac{\nabla^2\vartheta}{3} 
 \biggr)
  - \H\,\frac{\delta\varphi}{\bar\varphi'}
 \la{def_R_varphi}
 \;, \\ 
 \mathcal{R}^{ }_v
 & \equiv & 
 - \biggl( 
  h^{ }_\rmii{D} + \frac{\nabla^2\vartheta}{3} 
   \biggr)
 + \H \, (h - v)
 \;, \la{def_R_v} \\ 
 \mathcal{R}^{ }_\T
 & \equiv & 
 - \biggl( 
  h^{ }_\rmii{D} + \frac{\nabla^2\vartheta}{3} 
   \biggr)
 - \H\, \frac{\delta T}{T'}
 \;. \la{def_R_T}
\ea
We demonstrate in \se\ref{sss:combined_R} 
that the equations governing their evolutions
can be decoupled from the other perturbations, 
so that the essential part of 
${\mathcal{D}}\hspace*{0.4mm}'_{ }(\bar\Phi)$ 
reduces to a $3\times 3$ block. 

%
\subsection{Full system of equations in conformal time}
\la{sss:combined_R}

In order to implement the 
strategy outlined in \se\ref{sss:general}, we start 
with the equation for the inflaton field, from \se\ref{ss:inflaton}.
The goal is to convert this to an equation for the curvature 
perturbation $\mathcal{R}^{ }_\varphi$, from \eq\nr{def_R_varphi}.

When we insert $\varphi = \bar\varphi + \delta\varphi$ from 
\eq\nr{inflaton} into \eq\nr{varphi_eq}, the leading term yields
the background equation. Given that \eq\nr{def_R_varphi} includes
$\bar\varphi\hspace*{0.4mm}'$ and that a second time derivative
of~$\mathcal{R}^{ }_\varphi$ will appear, we are also encountered
with $\bar\varphi\hspace*{0.3mm}'''$,  
\ba
 && \bar\varphi\hspace*{0.3mm}'' 
 + \bigl( 2 \H  + a \Upsilon \bigr) \bar\varphi\hspace*{0.3mm}'
 + a^2 V^{ }_{\der\varphi} = 0 
 \la{bg_scalar_new} \\[2mm]
 & \Rightarrow & 
 \bar\varphi\hspace*{0.3mm}''' + \bigl[ 2 \H' + (a\Upsilon)' 
 - (2 \H + a \Upsilon)^2 \bigr] \bar\varphi\hspace*{0.3mm}' 
 + a^2 \bigl( V^{ }_{\der\varphi\varphi}\, \bar\varphi\hspace*{0.3mm}'
            + V^{ }_{\der\varphi\T}\, T'
            - a \Upsilon V^{ }_{\der\varphi} \bigr) 
 =
 0 
 \;. \la{bg_scalar_new_2} \hspace*{6mm}
\ea
For first-order perturbations, we make use of 
\eqs\nr{varphi_kinetic} and \nr{varphi_friction}, 
obtaining
\ba
 && \hspace*{-0.5cm}
 \delta\varphi{\hspace*{0.4mm}}''
 + \bigl(2\H + a \Upsilon\bigr)  \delta\varphi{\hspace*{0.4mm}}'
 - \nabla^2 \delta\varphi 
 + \bigl( a^2 V^{ }_{\der\varphi\varphi}  
           + a \bar\varphi' \Upsilon^{ }_{\der\varphi} \bigr) 
  \, \delta\varphi  
 \nn[2mm] 
 & &  
 + \,
  \bigl(
   2 a^2 V^{ }_{\der\varphi} 
 + a \Upsilon \bar\varphi'
 \bigr)\, h^{ }_0
 - \bar\varphi' \,
 \bigl( h_0' + 3 h_\rmii{D}' + \nabla^2 h \bigr)  
 + 
 \bigl( a^2 V^{ }_{\der\varphi\T} + a \bar\varphi' \Upsilon^{ }_{\der\T} \bigr)
 \, \delta T
 = 
 a^2_{ }\varrho 
 \;. \la{rescaled_pert_scalar_new}
 \hspace*{8mm}
\ea 

On the side of Einstein equations, 
we need the energy-momentum tensor from \se\ref{sss:Tmunu}. 
Anticipating again that the 2nd time derivative of $\H$ will
be needed for $\mathcal{R}_\varphi''$, 
the background equations read
\ba
 && \hspace*{-1cm}
 \H^2_{ } = \frac{4\pi G}{3}
 \bigl[ (\bar\varphi')^2_{ }  + 2 a^2 \bar{e}
 \bigr]
 \;, \quad 
 \H' = \frac{4\pi G}{3}
 \bigl[ -  2 (\bar\varphi')^2_{ } - a^2 (\bar{e} + 3 \bar{p})
 \bigr]
 \la{bg_einstein_new} 
 \\[2mm]
 & \Rightarrow & 
 \Biggl\{ 
 \begin{array}{l} 
 \H^2 - \H' \; = \; 4 \pi G \bigl[ (\bar\varphi')^2_{ }
                                   + a^2(\bar{e} + \bar{p})
                            \bigr]
 \\[2mm]
 2 \H^2 + \H' \; = \; 4 \pi G  a^2(\bar{e} - \bar{p})
 \end{array}
  \;,
 \la{bg_2} \\[2mm]
 & \Rightarrow & 
 \Biggl\{ 
 \begin{array}{l} 
 2\H\H' - \H'' \; = \; 4\pi G  \bigl\{ 
   2 \bar\varphi' \bar\varphi'' + 
  a^2 \bigl[ \bar{e}' + \bar{p}' + 2\H(\bar{e} + \bar{p}) \bigr] 
   \bigr\}
 \\[2mm]
 \H'' + 2 \H \H' - 4 \H^3 \; = \; 4 \pi G  a^2(\bar{e}' - \bar{p}') 
 \end{array}
 \;, \la{bg_einstein_new_2}
\ea
where $\bar{p}$ and $\bar{e}$ are from 
\eqs\nr{p_mixed} and \nr{e_mixed}, respectively. 
Sometimes it is also helpful to employ the background equation from 
${T^{\mu}_{ }}^{ }_{0;\mu} = 0$ even if it is not independent of 
the above, 
\be
 (\bar\varphi\hspace*{0.3mm}''
 + 2 \H \bar\varphi\hspace*{0.3mm}')\, \bar\varphi\hspace*{0.3mm}' 
 + 
 a^2 \bigl[  \bar{e}' + 3\H (\bar{e} + \bar{p} )\bigr]
 \; = \; 
 0
 \;.  \la{bg_Tmunu}
\ee

At first order in perturbations, 
considering the components 
${{G}^0_{ }}^{ }_0$ [$\Rightarrow$ \eq\nr{pert_einstein_00_new}], 
${{G}^0_{ }}^{ }_i$ [$\Rightarrow$ \eq\nr{pert_einstein_0i_new}], 
the traceless (or non-diagonal) part of 
${{G}^i_{ }}^{ }_j$ [$\Rightarrow$ \eq\nr{pert_einstein_ij_traceless_new}],
as well as  the trace part of 
${{G}^i_{ }}^{ }_j$ combined with ${{G}^0_{ }}^{ }_0$ and 
$\nabla^2/3$ times the traceless part of 
${{G}^i_{ }}^{ }_j$
[$\Rightarrow$ \eq\nr{from_einstein_sum_new}], 
and assuming that perturbations vanish at spatial
infinity so that $f^{ }_{,i} = 0$ implies $f=0$,
gives
\ba
 && \hspace*{-4.5cm}
 3 \H^2 h^{ }_0 
 + \H  \bigl( 3 h_\rmii{D}' + \nabla^2 h ) 
 - \nabla^2 \biggl( h^{ }_\rmii{D} + \frac{\nabla^2\vartheta}{3} \biggr) 
 \nn[2mm] 
 & = &  
  4 \pi G \Bigl[\,
  \bar\varphi\hspace*{0.4mm}' 
  \bigl(\, 
    h^{ }_0\, \bar\varphi'
 - \delta\varphi\hspace*{0.4mm}' 
   \,\bigr)
 - a^2_{ } \bigl(\, \bar{e}^{ }_{\der\T}\, \delta T 
                + \bar{e}^{ }_{\der\varphi}\, \delta\varphi \,\bigr) 
 \,\Bigr]
 \;, \la{pert_einstein_00_new} \hspace*{5mm} \\[2mm]
 \H h^{ }_0
 + \biggl( h^{ }_\rmii{D} + \frac{\nabla^2\vartheta}{3} \biggr)'
 & = & 
  4 \pi G \bigl[\, \bar\varphi'\, \delta\varphi
 - a^2(\bar{e} + \bar{p}) (h-v) \,\bigr]
 \;, \la{pert_einstein_0i_new} \\[2mm]
 && \hspace*{-4.5cm} 
 h^{ }_0 
 + \bigl(\partial^{ }_\tau + 2 \H \bigr) (h - \vartheta')
 - \biggl( h^{ }_\rmii{D} + \frac{\nabla^2\vartheta}{3} \biggr)
 \; = \; 
   0 
 \;, \la{pert_einstein_ij_traceless_new} \\[2mm]
 && \hspace*{-4.5cm}
 2 \bigl( \H' + 2 \H^2 \bigr) h^{ }_0 
 +
 \H \bigl( h_0' + 3 h_\rmii{D}' + \nabla^2 h \bigr)
 + \bigl(\partial^{2}_\tau + 2 \H \partial^{ }_\tau - \nabla^2 \bigr)
    \biggl( h^{ }_\rmii{D} + \frac{\nabla^2\vartheta}{3} \biggr)
 \nn[2mm] 
 & = &  
 4 \pi G a^2_{ } \bigl[\,
 (\bar{p}^{ }_{\der\T} - \bar{e}^{ }_{\der\T})\, \delta T 
 + 
 (\bar{p}^{ }_{\der\varphi} - \bar{e}^{ }_{\der\varphi})\, \delta\varphi
 \,\bigr] 
 \;. \la{from_einstein_sum_new}
\ea

With the help of 
\eqs\nr{pert_einstein_0i_new} and \nr{from_einstein_sum_new}, 
the metric perturbations 
$h^{ }_0$ and 
$h_0' + 3 h_\rmii{D}' + \nabla^2 h$
can be eliminated from \eq\nr{rescaled_pert_scalar_new}.
The result is conveniently expressed in terms of the object
\ba
 \frac{\H}{\bar\varphi\hspace*{0.3mm}'}
 \biggl( \frac{\bar\varphi\hspace*{0.3mm}'}{\H} \biggr)'_{ }
 \!\!\!
  & = &  
 \!\!
   \frac{ \bar\varphi\hspace*{0.3mm}''_{ } }
        { \bar\varphi\hspace*{0.3mm}'_{ } }
  - \frac{\H'}{\H} 
 \;. \la{pre_cal_F}
\ea
After a tedious reshuffling, and going over to  
the variables in \eqs\nr{def_R_varphi}--\nr{def_R_T}, 
we obtain 
\be
   \mathcal{M}^{ }_{1\varphi}\, \mathcal{R}^{ }_\varphi 
 + \mathcal{M}^{ }_{1v} 
   \bigl(\mathcal{R}^{ }_v - \mathcal{R}^{ }_\varphi\bigr)  
 + \mathcal{M}^{ }_{1\T} 
   \bigl(\mathcal{R}^{ }_\T - \mathcal{R}^{ }_\varphi\bigr)  
 \; = \; 
 - \frac{\varrho\, a^2_{ }\H}{ \bar\varphi\hspace*{0.3mm}' }
 \;,  \la{eq_R_varphi} 
\ee
where the coefficient operators read
\ba
 \mathcal{M}^{ }_{1\varphi}
 & = & 
   \partial_\tau^2 - \nabla^2_{ }
   + 
   \biggl[\, 
    a \Upsilon + 2 \H + 
    \frac{2\H}{\bar\varphi\hspace*{0.3mm}'}
    \biggl( \frac{\bar\varphi\hspace*{0.3mm}'}{\H} \biggr)'_{ }
   \,\biggr]\,\partial^{ }_\tau
  \;, \la{M_1varphi}
  \\[3mm]
 \mathcal{M}^{ }_{1v}
 & = &  -\, 
 \biggl[\, 
   a \Upsilon
   +
   \frac{2 \H}{\bar\varphi'}
   \biggl( \frac{\bar\varphi'}{\H} \biggr)'   
 \,\biggr]
    \frac{ 4\pi G a^2_{ }(\bar{e} + \bar{p}) }{\H }
  \;, 
  \\[3mm]
 \mathcal{M}^{ }_{1\T}
 & = &   
 \biggl[\, 
   \frac{4\pi G a^2_{ } ( \bar{e}^{ }_{\der\T}  - \bar{p}^{ }_{\der\T} ) }{ \H}
  +
  \frac{  
    a^2_{ } V^{ }_{\der\varphi\T}  
  +  a \bar\varphi\hspace*{0.3mm}' \Upsilon^{ }_{\der\T}  }
  { \bar\varphi\hspace*{0.3mm}' }
 \,\biggr] \, T\hspace*{0.3mm}' 
 \;.
\ea

We note that in the vacuum limit, i.e.\ $T\to 0$, the 
coefficients 
$
  \mathcal{M}^{ }_{1v}
$
and
$
 \mathcal{M}^{ }_{1\T}
$
vanish, because 
$\bar{e} + \bar{p}$ is proportional to 
the entropy density (cf.\ \eq\nr{e_plus_p}).
In this situation, other variables are often introduced,  
notably a gauge-invariant completion of $\delta\varphi$, 
$
 \mathcal{Q}^{ }_\varphi \equiv 
 - ({\bar\varphi\hspace*{0.3mm}'}/{\H})
 \, \mathcal{R}^{ }_\varphi
 = \delta\varphi + (\bar\varphi\hspace*{0.3mm}'/\H)
   (h^{ }_\rmiii{D} + \frac{\nabla^2\vartheta}{3})
$.
Its ``conformal'' version is  
$
 \widehat{\mathcal{Q}}^{ }_\varphi \equiv a \mathcal{Q}^{ }_\varphi
$.
Omitting $\Upsilon$ and $\varrho$, 
\eqs\nr{eq_R_varphi} and \nr{M_1varphi} yield a particularly 
simple equation for the latter, 
\be
   \biggl[\, 
      \partial_\tau^2 - \nabla^2_{ }
    - \frac{\H}{a\bar\varphi\hspace*{0.4mm}'}
      \biggl( \frac{a\bar\varphi\hspace*{0.4mm}'}{\H} \biggr)''_{ } 
   \,\biggr]
   \, \widehat{\mathcal{Q}}^{ }_\varphi
   \; 
    \overset{ \varrho,\Upsilon,T \,\to\, 0 }{  
    \underset{  }{ = } } 
   \;
   0
   \;. \la{eq_hatQ_varphi}
\ee
We will return to this when discussing 
initial conditions, around \eqs\nr{trafo_1}--\nr{eq_hatQ}.

\vspace*{3mm}

While all three combinations of perturbations appearing
in \eq\nr{eq_R_varphi}
are dimensionless and 
gauge invariant, the system is incomplete. 
Two further equations are needed for a unique specification 
of the solution (see below). 

Given that $\mathcal{M}^{ }_{1v}$ and $\mathcal{M}^{ }_{1\T}$
connect $\mathcal{R}^{ }_\varphi$ to 
plasma perturbations, it seems wise to search for
the additional equations from the same source that would be used 
for a purely hydrodynamic consideration, namely 
energy-momentum conservation, ${T^{\mu}_{ }}^{ }_{\nu;\mu} = 0$.\footnote{%
 If we wanted to proceed directly in terms of 
 \eqs\nr{pert_einstein_00_new}--\nr{from_einstein_sum_new}, 
 then \eqs\nr{pert_einstein_0i_new} and \nr{pert_einstein_ij_traceless_new}
 permit to express two gauge-invariant metric perturbations in terms
 of the variables in \eqs\nr{def_R_varphi}--\nr{def_R_T}, 
 cf.\ \eqs\nr{bardeen_1} and \nr{bardeen_2}. 
 The remaining two equations, 
 \eqs\nr{pert_einstein_00_new} and \nr{from_einstein_sum_new}, 
 contain the additional information needed for obtaining
 \eqs\nr{eq_R_v} and \nr{eq_R_T}. 
 However, implementing this in practice is extremely tedious.   
 } 
This is convenient also in the sense that the friction 
coefficient $\Upsilon$ does not appear explicitly. 
We recall that $\delta {T^{\mu}_{ }}^{ }_{\nu}$ is given 
in \eqs\nr{delta_Tmunu_varphi} and \nr{delta_Tmunu_r}.

A particularly simple equation originates from the $\nu = i$
components of energy-momentum conservation, {\it viz.}\ 
\be
 \bigl( \bar\varphi'' + 2 \H \bar\varphi'\bigr) \delta \varphi
 - 
 a^2_{ } \bigl\{ 
     \bar{p}^{ }_{\der\T} \delta T
   + \bar{p}^{ }_{\der\varphi} \delta\varphi
   + [(\bar{e} + \bar{p})(h-v)]'
   + (\bar{e} + \bar{p}) [ h^{ }_0 + 4 \H  (h-v) ]
 \bigr\}
 = 
 0
 \;. \la{dTmunu_i}
\ee
Inserting $h^{ }_0$ from \eq\nr{pert_einstein_0i_new}, 
making use of the background identities in 
\eqs\nr{bg_einstein_new} and \nr{bg_2}, 
and employing
the same variables as in \eq\nr{eq_R_varphi}, we get 
\be
   \mathcal{M}^{ }_{2\varphi}\, \mathcal{R}^{ }_\varphi 
 + \mathcal{M}^{ }_{2v} 
   \bigl(\mathcal{R}^{ }_v - \mathcal{R}^{ }_\varphi\bigr)  
 + \mathcal{M}^{ }_{2\T} 
   \bigl(\mathcal{R}^{ }_\T - \mathcal{R}^{ }_\varphi\bigr)  
 \; = \; 0
 \;, 
 \la{eq_R_v}
\ee
where
\be
 \mathcal{M}^{ }_{2\varphi}
  =
 \partial^{ }_\tau
 \;, \quad
 \mathcal{M}^{ }_{2v}
 = 
 \partial^{ }_\tau
 + 
    \frac{3 \H^2_{ }+ 4\pi G (\bar\varphi\hspace*{0.3mm}')^2_{ } }{\H }
    + \frac{(\bar e + \bar p)'}{\bar e + \bar p}
 \;, \quad
 \mathcal{M}^{ }_{2\T}
 = 
 - 
 \frac{\bar{p}^{ }_{\der\T} \, T\hspace*{0.3mm}'}{\bar e + \bar p} 
 \;.
\ee

A more complicated equation comes from 
the energy conservation part $\nu = 0$,
\ba
 && 
 \bar\varphi' \, 
 \bigl(\, \delta\varphi'' - \nabla^2 \delta\varphi \,\bigr)
 + 
 \bigl(\, \bar\varphi'' + 4\H \bar\varphi' \,\bigr)
 \delta\varphi'
 - 
 \bigl(\,
  h_0' + 3 h_\rmii{D}' + \nabla^2 h 
 \,\bigr) 
 (\bar\varphi' )^2_{ }
 - 2 h^{ }_0\, \bar\varphi' 
 \bigl(\, \bar\varphi'' + 2 \H \bar\varphi' \,\bigr)
 \nn[2mm]
 & + &  
 a^2\bigl[\, 
 \delta e' + 3\H \bigl( \delta e + \delta p \bigr)
 - \bigl( \bar{e} + \bar{p} \bigr)
   \bigl(\,
   \nabla^2 v + 3 h_\rmii{D}'  
   \,\bigr)
 \,\bigr]
 \; = \; 
 0
 \;. \la{dTmunu_0}
\ea
We note that for error-prone relations such as this one, 
it is advisable to crosscheck their gauge independence, by inserting
\eqs\nr{gauge1}--\nr{gauge7}. Furthermore, once we eliminate
variables, as described below, gauge independence can be 
rechecked after every step. 

Now, 
on the latter row of \eq\nr{dTmunu_0}, 
the velocity can be written as $v = h - (h-v)$, 
so that it appears in the same combination as in \eq\nr{def_R_v}.
Subsequently 
the metric perturbations 
$h^{ }_0$ and 
$h_0' + 3 h_\rmii{D}' + \nabla^2 h$
can be eliminated by employing 
\eqs\nr{pert_einstein_0i_new} and \nr{from_einstein_sum_new}, 
and the additional $h_0'$ by taking a time derivative
of \eq\nr{pert_einstein_0i_new}.
After quite some work, we find 
\be
   \mathcal{M}^{ }_{3\varphi}\, \mathcal{R}^{ }_\varphi 
 + \mathcal{M}^{ }_{3v} 
   \bigl(\mathcal{R}^{ }_v - \mathcal{R}^{ }_\varphi\bigr)  
 + \mathcal{M}^{ }_{3\T} 
   \bigl(\mathcal{R}^{ }_\T - \mathcal{R}^{ }_\varphi\bigr)  
 \; = \; 0
 \;, 
 \la{eq_R_T}
\ee
where
\ba
 \mathcal{M}^{ }_{3\varphi}
 & = & -
 \frac{(\bar\varphi\hspace*{0.3mm}')^2_{ }}{a^2_{ }}
 \biggl\{\, 
   \partial_\tau^2 - \nabla^2_{ }
   + 
   2 \biggl[\, 
    \H + 
    \frac{\H}{\bar\varphi\hspace*{0.3mm}'}
    \biggl( \frac{\bar\varphi\hspace*{0.3mm}'}{\H} \biggr)'_{ }
   \,\biggr]\,\partial^{ }_\tau
  \,\biggr\}
  \nn[2mm]
 & + &
 \bigl( \bar{e} + \bar{p} \bigr) 
 \,\biggl[\,
   \biggl( 2\H + \frac{\H'}{\H} \biggr)
   \,\partial^{ }_\tau + \nabla^2_{ }
 \,\biggr]
 \;, \\[3mm]
 \mathcal{M}^{ }_{3v}
 & = &
 + \frac{(\bar\varphi\hspace*{0.3mm}')^2_{ }}{a^2_{ }}
 \biggl\{\, 
    \frac{2 \H}{\bar\varphi\hspace*{0.3mm}'}
    \biggl( \frac{\bar\varphi\hspace*{0.3mm}'}{\H} \biggr)'_{ }
    \frac{4\pi G a^2(\bar{e} + \bar{p})}{\H}
 \biggr\}
 \nn[2mm]
 & - &
 \bigl( \bar{e} + \bar{p} \bigr) 
 \biggl\{\, 
    \frac{4\pi G a^2(\bar{e} + \bar{p})}{\H}
 \biggl[\,
    \partial^{ }_\tau
    + 6 \H 
    + \frac{(\bar e + \bar p)'}{\bar e + \bar p}   
 \,\biggr]
 -\nabla^2_{ }
 \,\biggr\} 
 \;, 
 \\[3mm]
 \mathcal{M}^{ }_{3\T}
 & = &
 - \bar{e}^{ }_{\der\T}  \, T\hspace*{0.3mm}'
 \biggl\{\,
   \partial^{ }_\tau 
  + 2 \, \biggl(2\H - \frac{\H'}{\H} \biggr)
  + \frac{(\bar{e}^{ }_{\der\T} T\hspace*{0.3mm}')\hspace*{0.3mm}'}
         {\bar{e}^{ }_{\der\T} T\hspace*{0.3mm}'}
  + \biggl(2\H + \frac{\H'}{\H} \biggr)
    \frac{\bar{p}^{ }_{\der\T}}{\bar{e}^{ }_{\der\T}}
 \,\biggr\}
 \;. \la{calM_3T}
\ea
Together with \eqs\nr{eq_R_varphi} and \nr{eq_R_v}, this
completes the system needed for determining the origin and 
evolution of the three gauge-invariant perturbations.

%
\subsection{Full system of equations in physical time}
\la{ss:physical_time}

In \se\ref{sss:combined_R}, we have derived the basic equations
in conformal time, as this is frequently convenient for theoretical
considerations. However, for a practical numerical solution, it is 
advantageous to make use of physical time. We show that the set of 
equations simplifies remarkably after this change of variables. 
 
In physical time, 
the background identities, from 
\eqs\nr{bg_scalar_new}, \nr{bg_einstein_new} and \nr{bg_Tmunu}, 
become 
\be
  \ddot{\bar\varphi} 
 \, = \, 
 - \,(3 H + \Upsilon)\,\dot{\bar\varphi} - V^{ }_{\der\varphi}
 \;, 
 \quad 
 \dot{H} 
 \, = \, 
 -\,4\pi G \,\bigl(\, 
 \dot{\bar\varphi}^2_{ } + \bar{e} + \bar{p}
 \,\bigr)
 \;, 
 \quad 
 3 H^2_{ } 
 \, = \, 
  4\pi G \,(\, \dot{\bar\varphi}^2 + 2 \bar{e}\, )
 \;,  \la{bg_t} 
\ee
\be
 (\ddot{\bar\varphi} + 3 H \dot{\bar\varphi})\, \dot{\bar\varphi}
 + 
 \bigl[  \dot{\bar{e}} + 3 H (\bar{e} + \bar{p} )\bigr]
 \; = \; 
 0
 \;.  \la{bg_Tmunu_new}
\ee
These are re-expressed in a form suitable for practical
integration in \eqs\nr{eom_field}--\nr{eom_friedmann}.

As for the perturbations, 
the coefficient function form 
\eq\nr{pre_cal_F} is now represented by
\ba
 \hcoeff & \equiv & 
 \frac{H}{\dot{\bar\varphi}}
 \biggl( \frac{\dot{\bar\varphi}}{H} \biggr)^{\textstyle .}_{ }   
 \; = \; 
 \frac{\ddot{\bar\varphi}}{\dot{\bar\varphi}} 
 - \frac{\dot H}{H}  
 \;. \la{cal_F} 
\ea
We note that in the slow-roll regime, where $H\simeq$ constant and 
$\ddot{\bar\varphi}$ is small, $|\hcoeff| \ll H$.

Inspecting the operators $\mathcal{M}^{ }_{iv}$ and $\mathcal{M}^{ }_{i\T}$, 
$i\in\{1,2,3\}$,
it is apparent that the equations can be simplified by rescaling the original
variables $\mathcal{R}^{ }_v - \mathcal{R}^{ }_\varphi$ and 
$\mathcal{R}^{ }_\T - \mathcal{R}^{ }_\varphi$.
The optimal set\hspace*{0.3mm}\footnote{%
 Here ``optimal'' refers to the appearance of the equations.
 However, at late times in a radiation-dominated universe, when
 $t \sim \mpl^{ }/T^2_{ }$ and $\dot{T} \sim - T^3_{ } / \mpl^{ }$, 
 the prefactors 
 $
  \bar{e} + \bar{p}
 $ and 
 $
   \bar{e}^{ }_{\der\T}\,\dot{T}
 $
 decrease as $T^4_{ }$ and $T^6_{ }/\mpl$, 
 respectively. In that regime, it is more efficient 
 to adopt $\mathcal{R}^{ }_v$ and $\mathcal{R}^{ }_\T$ as 
 variables, as we will do in \se\ref{ss:radiation}, 
 whereas $\mathcal{R}^{ }_\varphi$ decouples
 from the evolution. \la{scaling}
 }
now consists of
\be
 \mathcal{R}^{ }_\varphi 
 \;, \quad
 \E^{ }_v \; \equiv \;  
 (\bar{e} + \bar{p}) (\mathcal{R}^{ }_v - \mathcal{R}^{ }_\varphi)
 \;, \quad
 \E^{ }_\T \; \equiv \;  
 \bar{e}^{ }_{\der\T}\, \dot{T}\,
 (\mathcal{R}^{ }_\T - \mathcal{R}^{ }_\varphi)
 \;. \la{optimal_set}
\ee
We remark that both $\E^{ }_v$ 
and $\E^{ }_\T$ can have a non-zero value
only in the presence of non-vanishing temperature
and entropy density,
because \eq\nr{e_mixed} implies that 
$\bar e^{ }_{\der\T} = T \bar s^{ }_{\der\T}$
and \eq\nr{e_plus_p} that
$\bar e + \bar p = T \bar s$.
Since the ``curvature'' part, from 
\eq\nr{def_R_x}, cancels in the differences 
defining $\E^{ }_v$ and 
$\E^{ }_\T$, we could call these  
``isocurvature'' perturbations. The latter is also 
proportional to what is called the total entropy 
perturbation, cf.\ \eq\nr{entropy_pert}.

Assuming translational invariance in spatial directions, we
go over to comoving momentum space, with $-\nabla^2_{ }\to k^2_{ }$.
Employing the background identities from 
\eq\nr{bg_t},
we then find that \eqs\nr{eq_R_varphi},  
\nr{eq_R_v} and \nr{eq_R_T} can be converted into 
\ba
 \ddot{\mathcal{R}}^{ }_\varphi
 & = & 
 - \frac{\varrho H}{\dot{\bar\varphi}}
 - \dot{\mathcal{R}}^{ }_\varphi 
    \, \bigl[\, \Upsilon + 3 H + 2 \hcoeff \,\bigr]
 -  \mathcal{R}^{ }_\varphi 
    \, \biggl[\, \frac{k^2_{ }}{a^2_{ }} \,\biggr]
 \nn[2mm]
 & +  & 
   \E^{ }_v\, 
   \biggl[\,
       \frac{4\pi G ( \Upsilon  +  2 \hcoeff )}{H}
   \,\biggr] 
 - \E^{ }_\T \, 
   \biggl[\,
   \frac{4\pi G}{H}
 \biggl( 1 - \frac{\bar{p}^{ }_{\der\T}}{\bar{e}^{ }_{\der\T}}\biggr)
   + \frac{V^{ }_{\der\varphi\T} 
   + \Upsilon^{ }_{\der\T} \dot{\bar\varphi} }
     {\dot{\bar\varphi} \, \bar{e}^{ }_{\der\T}}
   \,\biggr]
 \;, \hspace*{5mm} \la{dot_R_varphi}
 \\[3mm]
 \dot{\E}^{ }_v
 & = & 
  - \dot{\mathcal{R}}^{ }_\varphi 
    \, \bigl[ \bar{e} + \bar{p} \bigr]
  - \E^{ }_v 
 \biggl[\, 
 3 H + 
 \frac{4 \pi G \dot{\bar\varphi}^2_{ }}{H}
 \,\biggr]
 + 
 \E^{ }_\T
 \,
 \biggl[\, 
   \frac{\bar{p}^{ }_{\der\T}}{\bar{e}^{ }_{\der\T}}
 \,\biggr]
 \;, \la{dot_S_v}
 \\[3mm]
 \dot{\E}^{ }_\T
 & = & 
 \varrho \, \dot{\bar\varphi}  H
 + \dot{\mathcal{R}}^{ }_\varphi\,
  \biggl[\,
   \Upsilon\dot{\bar\varphi}^2_{ }
  + \frac{8\pi G \bar{e}(\bar{e}+\bar{p})}{H}
  \,\biggr]
 - \mathcal{R}^{ }_\varphi \, 
   \biggl[\, 
     (\bar{e} + \bar{p}) \frac{k^2_{ }}{a^2_{ }}
   \,\biggr]
 \nn[2mm]
 & - & 
 \E^{ }_v \,
 \biggl[\,
  \frac{k^2_{ }}{a^2_{ }} 
 + \frac{4\pi G }{H}
   \biggl(\, \Upsilon \dot{\bar\varphi}^2_{ } 
       + \frac{8 \pi G \bar{e}(\bar{e} + \bar{p})}{H}
   \,\biggr)
 \,\biggr]
 \nn[2mm]
 & + & 
 \E^{ }_\T \, 
 \biggl[\,
   \frac{\dot{H} - 4\pi G(\bar{e} + \bar{p})}{H}
  - 3 H \,\biggl( 1 
  + \frac{\bar{p}^{ }_{\der\T}}{\bar{e}^{ }_{\der\T}}\,\biggr)
   + \frac{( V^{ }_{\der\varphi\T} 
   + \Upsilon^{ }_{\der\T} \dot{\bar\varphi})\,\dot{\bar\varphi}  }
     {\bar{e}^{ }_{\der\T}}
 \,\biggr]
 \;. \la{dot_S_T}
\ea
These equations constitute our main result. 

Equations~\nr{dot_R_varphi}--\nr{dot_S_T} 
possess an important property, 
which becomes apparent if we look for a stationary solution, 
$
 \ddot{\mathcal{R}}^{ }_\varphi 
 = 
 \dot{\mathcal{R}}^{ }_\varphi
 = 
 \dot{\E}^{ }_v
 = 
 \dot{\E}^{ }_\T
 = 
 0
$.
It originates from the fact that in this limit, 
the coefficient of each appearance of 
$\mathcal{R}^{ }_\varphi$ is proportional to $k^2_{ }/a^2_{ }$.
If we put $k/a\to 0$, or more physically $k/a \ll H$,
so that we are outside of the Hubble horizon, 
and omit the noise~$\varrho$, we obtain a system of three
linear equations for the two perturbations 
$ {\E}^{ }_v $
and
$ {\E}^{ }_\T $.
The coefficients in these equations 
are non-degenerate. Therefore the system 
is overconstrained, leading to the only solution 
$\E^{ }_v = \E^{ }_\T = 0$. 
Then, according to \eq\nr{optimal_set}, all curvature perturbations agree, 
$\mathcal{R}^{ }_v = \mathcal{R}^{ }_\T = \mathcal{R}^{ }_\varphi$.
This conforms with the general proof in ref.~\cite{weinberg},
where curvature perturbations were called adiabatic ones, and
non-adiabatic perturbations were shown to vanish
in single-field inflation.

%
\subsection{Other variables from the literature}
\la{ss:other}

We explained in \se\ref{sss:general} that our system 
contains five independent gauge-invariant variables. 
However, in \eqs\nr{def_R_varphi}--\nr{def_R_T}, only three variables
have been defined, whose time evolutions are dictated by
\eqs\nr{dot_R_varphi}--\nr{dot_S_T}. 
Here we show how other 
gauge-invariant variables can be obtained, 
once the solutions for 
$\mathcal{R}^{ }_\varphi$,
$\E^{ }_v$ and
$\E^{ }_\T$ 
(and subsequently $\mathcal{R}^{ }_v$ and
$\mathcal{R}^{ }_\T$) are known. 

Two further gauge-invariant variables 
can be chosen as the Bardeen potentials~\cite{bardeen}. 
There are a plethora of notations for the Bardeen potentials
in the literature; we adopt 
\ba
 \phi & \equiv & 
 h^{ }_0 + \bigl(\partial^{ }_\tau + \H \bigr)
 \bigl(\, h -\vartheta' \,\bigr)
 \;, \la{bd_phi} \\[2mm] 
 \psi & \equiv & 
 h^{ }_\rmii{D} + \frac{\nabla^2_{ }\vartheta}{3}
 - \H \, \bigl(\, h - \vartheta' \,\bigr)
 \;. \la{bd_psi} 
\ea
One equation for these can be obtained from 
\eq\nr{pert_einstein_ij_traceless_new}, expressing the left-hand
side in terms of $\phi$ and $\psi$,
\ba
 \phi - \psi & = & 0 
 \;. \la{bardeen_1}
\ea 
The other equation can 
be obtained from \eq\nr{pert_einstein_0i_new}.
To do this, we first express the right-hand side in terms of
$\mathcal{R}^{ }_\varphi$ and $\mathcal{R}^{ }_v$, via
$
 \delta\varphi = 
 -
 [
 \mathcal{R}^{ }_\varphi 
 + (h^{ }_\rmiii{D} + \frac{\nabla^2\vartheta}{3})
 ] ( \bar\varphi\hspace*{0.4mm}' / \H )
$
and
$
 (h-v) 
 = 
 [
 \mathcal{R}^{ }_v + (h^{ }_\rmiii{D} + \frac{\nabla^2\vartheta}{3})
 ]/\H
$.
Moving the metric perturbations to the left-hand side, 
going over to physical time,
and making use of a background identity from \eq\nr{bg_t},  
\eq\nr{pert_einstein_0i_new} becomes
\ba
 H \phi + \biggl( \partial^{ }_t - \frac{\dot{H}}{H} \biggr) \psi
 & = & 
 -\, 
 \frac{4\pi G}{H}
 \bigl[\, 
    \dot{\bar\varphi}^2_{ } \,\mathcal{R}^{ }_\varphi
 + 
   (\bar{e} + \bar{p}) \,\mathcal{R}^{ }_v
 \,\bigr] 
 \;. \la{bardeen_2}
\ea
It is remarkable that even though 
the Bardeen potentials are often 
introduced as the key variables of the problem, they have turned into 
auxiliary quantities in our setup, not influencing the evolution
of $\mathcal{R}^{ }_\varphi$, $\mathcal{R}^{ }_v$ and $\mathcal{R}^{ }_\T$ 
(though sourced by the first two of them). 

\vspace*{3mm}

Yet another set of gauge-invariant variables consists of energy
density perturbations. Again there are many ways to define 
gauge-invariant linear combinations. As an example,  
with $\delta e$ from \eq\nr{e_perturbations},
making use of the variables in \eq\nr{optimal_set},
we denote
\ba
 \Delta^{ }_v
 & \equiv & 
 \frac{ 
  \delta e + \bar e \hspace*{0.3mm}' 
 \, (h-v) 
 }{ - \overline{T}\hspace*{0.3mm}{}^{0}_{ }{}^{ }_{0} }
 \; = \; 
 \frac{
   \bar{e}^{ }_{\der\T}\, \dot{T}\,
   (\mathcal{R}^{ }_v - \mathcal{R}^{ }_\T) 
 + \bar{e}^{ }_{\der\varphi}\, \dot{\bar\varphi}\, 
   (\mathcal{R}^{ }_v - \mathcal{R}^{ }_\varphi) 
 }{\bigl( \bar{e} + \tfr12 \dot{\bar\varphi}^2_{ } \bigr)\, H}
 \la{Delta_v}
 \;. 
\ea
We remark that our definition
of $\bar{e}$ from \nr{e_mixed}, and consequently
$\delta e$ and $\bar{e}\hspace*{0.2mm}'$ from \eq\nr{e_perturbations},  
include {\em no} time derivatives of~$\bar\varphi$.
Therefore $\bar{e}$ and $\delta e$
represent the full physical energy density and its
perturbations only once inflaton oscillations 
have attenuated.

Finally, the (total)
entropy perturbation is conventionally defined as
\be
 \mathcal{S}^{ }_\rmi{tot} \;\equiv\;
 \H\,\biggl( 
   \frac{\delta p}{\bar p\hspace*{0.3mm}'}
  -
   \frac{\delta e}{\bar e\hspace*{0.3mm}'}
 \biggr)
 \;. \la{entropy_pert}
\ee
If only one type of degree of freedom is active, for instance
either $\delta\varphi$ or $\delta T$ but not both at the same time, 
then it follows from \eqs\nr{p_perturbations} and \nr{e_perturbations} 
that $\mathcal{S}^{ }_\rmi{tot} = 0$. 
Indeed, in post-inflationary cosmology, entropy 
perturbations require the presence of several {\em different} 
matter components. 
The simplest way 
to generate an entropy perturbation is that 
$\delta\varphi$ and $\delta T$ exist as
simultaneous degrees of freedom.  
This becomes explicit if we
insert \eqs\nr{p_perturbations} and \nr{e_perturbations}, 
obtaining 
\be
 \mathcal{S}^{ }_\rmi{tot}
 \; = \; 
 \frac{
       (\, 
       \bar{e}^{ }_{\der\T}\,
       \bar{p}^{ }_{\der\varphi}
       -
       \bar{p}^{ }_{\der\T}\,
       \bar{e}^{ }_{\der\varphi}
       \,)\, \dot{\bar\varphi} \, \dot{T} 
      }{
       \dot{\bar e}\, \dot{\bar p}
      }
      \,
      \bigl(\,
        \mathcal{R}^{ }_\T - \mathcal{R}^{ }_\varphi 
      \,\bigr)
 \;. 
\ee

%
\section{How to complete the setup}
\la{se:subtleties}

The evolution equations \nr{dot_R_varphi}--\nr{dot_S_T} look
relatively simple,  but they are not analytically solvable. 
Indeed their solution is a 
complicated function, displaying different regimes, with oscillations in 
one or the other variable. In this section, we elaborate upon 
issues that need to be taken care of, before 
the equations can be efficiently solved numerically.

%
\subsection{Fixing initial conditions}
\la{ss:initial}

As \eqs\nr{dot_R_varphi}--\nr{dot_S_T} represent a linear system of 
equations, any solution could be multiplied by an overall constant,
and it would still represent a solution. That said, the relations 
between the components of the solution ``vector'',  
$
 \{\dot{\mathcal{R}}^{ }_\varphi,\mathcal{R}^{ }_\varphi,
   \E^{ }_v,\E^{ }_\T\}
$, 
adjust themselves dynamically to the correct values. 
The important thing is to fix the initial 
absolute magnitude of the 
solution vector. This is best done by going to a regime in which
one component dominates. Then we need to know precisely 
only the largest component of the solution vector.   

Let us consider \eq\nr{dot_R_varphi} at early times
when the energy density carried by the inflaton field far exceeds that
in the plasma, 
$
 \tfr{1}{2} \dot{\bar\varphi}^2_{ } + V - T V^{ }_{\der\T} \gg e^{ }_r
$.
Then it follows from \eq\nr{bg_t} that 
$
 4\pi G e^{ }_r \ll H^2_{ }
$.
The same holds for quantities in which the vacuum part cancels, 
notably $\bar{e} + \bar{p}$ as given by \eq\nr{e_plus_p}, whereby
$
 4\pi G (\bar{e} + \bar{p}) \ll H^2_{ }
$.

Consider now the contributions of $\E^{ }_v$ and $\E^{ }_\T$
to \eq\nr{dot_R_varphi}. We find 
$
 |\E^{ }_v| = (\bar{e}+\bar{p})|\mathcal{R}^{ }_v - \mathcal{R}^{ }_\varphi|
 \ll H^2_{ }|\mathcal{R}^{ }_v - \mathcal{R}^{ }_\varphi|/(4\pi G)
$, 
and therefore
\be
 \biggl| 
  \E^{ }_v\, 
   \biggl[\,
       \frac{4\pi G ( \Upsilon  +  2 \hcoeff )}{H}
   \,\biggr]
 \biggr|
 \ll 
 \bigl|\, \mathcal{R}^{ }_v - \mathcal{R}^{ }_\varphi \,\bigr|
 \,H\, \bigl|\, \Upsilon + 2 \hcoeff \,\bigr|
 \;. \la{est_1}
\ee
As for $\E^{ }_\T$, we note that the temperature settles to a value
at which the energy released from the inflaton compensates against 
that lost to Hubble expansion. Then, from \eq\nr{eom_plasma}, omitting
terms involving $V^{ }_{\der\T}$, we find
$
 |\bar{e}^{ }_{\der\T} \dot{T}| \ll 3 H (\bar{e} + \bar{p})
 \ll H^3_{ }/(4\pi G)
$.
This implies
\be
 \biggl| \E^{ }_\T \, 
   \frac{4\pi G}{H}
 \biggr| 
 \ll 
 \bigl|\, \mathcal{R}^{ }_\T - \mathcal{R}^{ }_\varphi \,\bigr|
 H^2_{ }
 \;, \quad
 \biggl| 
 \E^{ }_\T \, 
     \frac{ \Upsilon^{ }_{\der\T}  }
     { \bar{e}^{ }_{\der\T}}
 \biggr|
 \ll 
 \bigl|\, \mathcal{R}^{ }_\T - \mathcal{R}^{ }_\varphi \,\bigr| H \Upsilon
 \;, \la{est_2}
\ee 
where in the last case we assumed 
$\Upsilon^{ }_{\der\T}/\bar{e}^{ }_{\der\T} \sim \Upsilon / e^{ }_r$.

If we are in the regime $k/a \gg \{ H,\Upsilon \}$;  
remember that $\hcoeff \ll H$ at early times; 
and furthermore assume that 
$
 \mathcal{R}^{ }_\varphi \sim \mathcal{R}^{ }_v \sim \mathcal{R}^{ }_\T
$, 
then \eqs\nr{est_1} and \nr{est_2} imply that the contributions 
of $\E^{ }_v$ and $\E^{ }_\T$ can be omitted in \eq\nr{dot_R_varphi}. 
Therefore, $\mathcal{R}^{ }_\varphi$ evolves independently of 
$\E^{ }_v$ and $\E^{ }_\T$. We have verified numerically as well that
this is an excellent approximation 
for the benchmarks that we 
introduce in \se\ref{se:numerics}.  

We still need to determine the initial conditions for 
$\mathcal{R}^{ }_\varphi$. For this, it is helpful to rescale
variables like in \eq\nr{eq_hatQ_varphi}, 
\be
   \mathcal{Q}^{ }_\varphi     
   \; \equiv \; 
 - 
   \frac{\dot{\bar\varphi}}{H} 
 \, 
    \mathcal{R}^{ }_\varphi    
 \;, \quad
 \widehat{\mathcal{Q}}^{ }_\varphi 
 \; \equiv \; 
 a \mathcal{Q}^{ }_\varphi
 \;. \la{relation}
\ee
In the first step, 
\ba
 \dot{\mathcal{R}}^{ }_\varphi & = & 
 - \frac{H}{\dot{\bar\varphi}} 
 \,\biggl[\,
    \dot{\mathcal{Q}}^{ }_\varphi
  - 
    \frac{H}{\dot{\bar\varphi}}
 \biggl( \frac{\dot{\bar\varphi}}{H} \biggr)^{\textstyle .}_{ }   
    \mathcal{Q}^{ }_\varphi
 \,\biggr]
 \;, \la{trafo_1} \\[2mm]
 \ddot{\mathcal{R}}^{ }_\varphi
 + \frac{2 H}{\dot{\bar\varphi}}
 \biggl( \frac{\dot{\bar\varphi}}{H} \biggr)^{\textstyle .}_{ }
 \, \dot{\mathcal{R}}^{ }_\varphi
 & = & 
 - \frac{H}{\dot{\bar\varphi}} 
 \,\biggl[\,
    \ddot{\mathcal{Q}}^{ }_\varphi
  - 
    \frac{H}{\dot{\bar\varphi}}
 \biggl( \frac{\dot{\bar\varphi}}{H} \biggr)^{\textstyle ..}_{ }   
    \mathcal{Q}^{ }_\varphi
 \,\biggr]
 \;, \la{trafo_2}
\ea
and in the second, 
\ba
 \dot{\mathcal{Q}}^{ }_\varphi
 & = & 
 \frac{1}{a}
 \Bigl[\,
    \!\dot{\widehat{\mathcal{Q}}\,}^{ }_\varphi - 
   H\, \widehat{\mathcal{Q}}^{ }_\varphi
 \,\Bigr]
 \;, \la{trafo_3} \\[2mm]
 \ddot{\mathcal{Q}}^{ }_\varphi
  + 2 H \dot{\mathcal{Q}}^{ }_\varphi
 & = & 
 \frac{1}{a}
 \Bigl[\,
    \!\ddot{\widehat{\mathcal{Q}}\,}^{ }_\varphi - 
   \frac{\ddot{a}}{a}\, \widehat{\mathcal{Q}}^{ }_\varphi
 \,\Bigr]
 \;. \la{trafo_4}
\ea
Multiplying subsequently the first line of \eq\nr{dot_R_varphi} by 
$-a \dot{\bar\varphi}/H$, and returning to conformal time, 
we get a second-order differential 
equation for $\widehat{\mathcal{Q}}^{ }_\varphi$,
\be
   \biggl\{\, 
      \partial_\tau^2
    + k^2_{ }
    - \frac{\H}{a\bar\varphi\hspace*{0.4mm}'}
      \biggl( \frac{a\bar\varphi\hspace*{0.4mm}'}{\H} \biggr)''_{ } 
    +  a \Upsilon \biggl[\, 
    \partial^{ }_\tau 
    - \frac{\H}{a\bar\varphi\hspace*{0.4mm}'}
      \biggl( \frac{a\bar\varphi\hspace*{0.4mm}'}{\H} \biggr)'_{ } 
   \,\biggr]
   \,\biggr\}
   \, \widehat{\mathcal{Q}}^{ }_\varphi
   \; 
    \overset{ \E^{ }_v\,,\,\E^{ }_\T \,\to\, 0 }{  
    \underset{  }{ = } } 
   \;
   a^3_{ }\varrho
   \;. \la{eq_hatQ}
\ee

Equation~\nr{eq_hatQ} has an important implication. The first three
terms are those of a harmonic oscillator with a time-dependent mass
squared. 
If $\Upsilon > 0$, 
or more physically $\Upsilon > H$, 
the terms multiplied by $a\Upsilon$ play a role. 
Then the harmonic oscillator is damped. This means that any initial
condition that we may set gets rapidly attenuated by the damping, while 
simultaneously new fluctuations are generated by the noise term. 
This is the characteristic dynamics of the strong regime of warm inflation.

In contrast, if $\Upsilon < H$, the oscillations are only weakly damped.
Omitting $\Upsilon$ and $\varrho$ altogether, 
the equation of motion has the same form as in vacuum. 
Its normalization can be fixed by considering the quantization of 
a free vacuum scalar field. As is well known, 
with a time-dependent mass, the definition
of a vacuum state is not unique. However, 
for momenta large enough that
$k^2_{ }$ dominates, we are effectively in Minkowskian spacetime, 
and the problem is solvable. A brief overview on this can be 
found in appendix~\ref{se:quantum}.

We denote the Fourier transform of the inflaton fluctuation,
or more precisely the mode function, by $\delta \varphi^{ }_k$, 
recalling that it is complex-valued. 
A dimensionless version thereof is defined as 
$\delta\widehat{\varphi}^{ }_k \equiv a\, \delta \varphi^{ }_k$.
We treat $\delta\widehat{\varphi}^{ }_k$ as interchangeable to  
$\widehat{\mathcal{Q}}^{ }_\varphi$, even though the former
is gauge dependent and therefore a precise equivalence only
holds in a specific gauge; 
gauge dependence only affects the mass squared, 
whereas for initial conditions 
we are concerned with the much larger $k^2_{ }$
(cf.\ appendix~\ref{se:quantum} for a more detailed discussion).

The information needed for fixing the normalization 
of $\widehat{\mathcal{Q}}^{ }_\varphi$ 
can be obtained by going to early times, when $k/a \gg \{ \bar{m},H \}$, 
where $\bar{m}^2_{ } \equiv V^{ }_{\der\varphi\varphi}$.
In conformal time, this corresponds to $k|\tau| \gg 1$.
The forward-propagating solution, normalized to satisfy 
the canonical commutation relation, then reads (cf.\ \eq\nr{ex_mink})
\be
 \widehat{\mathcal{Q}}^{ }_\varphi 
 \; \stackrel{\frac{k}{a}\,\gg\, \bar{m},H}{\approx} \;
 \frac{e^{-i k \tau}_{ }}{\sqrt{2k}} 
 \;, \quad
 \,\dot{\!\widehat{\mathcal{Q}}}^{ }_\varphi 
 \; \stackrel{\frac{k}{a}\,\gg\, \bar{m},H}{\approx} \;
 \frac{e^{-i k \tau}_{ }}{\sqrt{2k}}
 \biggl( - \frac{ik}{a} \biggr)
 \;. \la{initial_dot_varphi}
\ee
An overall phase factor could be added, but plays no physical role. 

Given that the system of equations is linear, 
we may multiply all components of the solution vector
with a common constant. Since physics is contained in power spectra
of the type
\be
 \P^{ }_{\mathcal{O}^{ }_k}
 \equiv  
 \frac{k^3_{ }}{2\pi^2_{ }}
 \,\langle\,
   \mathcal{O}^{*}_k\, \mathcal{O}^{ }_k
 \,\rangle
 \;, \la{rescale}
\ee
a convenient choice is to carry out a rescaling with 
$
 k^{3/2}_{ } / (\sqrt{2}\pi)
$.
Ignoring the time evolution of ${H} / (a{\dot{\bar\varphi}})$
as this is slow compared with $k/a$,  
then suggests the initial conditions
(imposed at a time $t = t^{ }_1$, 
with $a(t^{ }_1) \equiv a^{ }_1$)
\ba
 \bigl[\, \dot{\mathcal{R}}^{ }_\varphi \,\bigr]^{ }_\rmi{rescaled} (t^{ }_1)
 & \simeq & 
 \frac{H}{\dot{\bar\varphi}} 
 \frac{  i }{2\pi}\, \frac{k^2_{ }}{a^2_{1}}
 \;, \la{initial_dot_R_varphi} \\[2mm] 
 \bigl[\, \mathcal{R}^{ }_\varphi \,\bigr]^{ }_\rmi{rescaled}(t^{ }_1)
 & \simeq & 
 - \frac{H}{\dot{\bar\varphi}} 
 \frac{ 1 }{2\pi}\, \frac{k}{a^{ }_1} 
 \;. \la{initial_R_varphi} 
\ea

A few remarks are in order. 
It is natural to impose the 
initial conditions during the attractor stage of inflation, so that 
all background quantities have settled into slow evolution.
Even though 
\eqs\nr{initial_dot_R_varphi} and \nr{initial_R_varphi} set
$\mathcal{R}^{ }_\varphi$ in rapid motion, 
this initial motion is a pure phase rotation, 
with $\partial^{ }_t |\mathcal{R}^{ }_\varphi|^2_{t = t^{ }_1} = 0$. 
Until the oscillation regime
(cf.\ \se\ref{ss:singularities}), 
the real and imaginary parts of~$\mathcal{R}^{ }_\varphi$ 
evolve independently of each other. 
Given that all
perturbations are coupled to each other
via \eqs\nr{dot_R_varphi}--\nr{dot_S_T}, 
$\E^{ }_v$ 
and $\E^{ }_\T$ need to be complexified as well. 
For $\E^{ }_v$ and $\E^{ }_\T$, 
we can start from vanishing initial conditions,
with $\mathcal{R}^{ }_\varphi \neq 0$ then 
rapidly sourcing non-zero values for them. 

To be convinced about the robustness of 
the initial conditions, 
we may carry out practical tests on variations thereof. 
If we account for the time derivative of $H/(a \dot{\bar\varphi})$ when 
relating $\dot{\mathcal{R}}^{ }_\varphi$ to 
$\,\dot{\!\widehat{\mathcal{Q}}}^{ }_\varphi$, 
then we are led to replace \eq\nr{initial_dot_R_varphi} through
\be
 \bigl[\, \dot{\mathcal{R}}^{ }_\varphi \,\bigr]^\rmi{alt}_\rmi{rescaled}
 (t^{ }_1)
 \; \simeq \; 
 \frac{H}{\dot{\bar\varphi}} 
 \,\biggl[\,
 \frac{  i }{2\pi}\, \frac{k^2_{ }}{a^2_{1}}
 +  
 \bigl(\, H + \hcoeff \,\bigr)\,
 \frac{ 1 }{2\pi}\, \frac{k}{a^{ }_1} 
 \,\biggr]
 \;, \la{initial_dot_R_varphi_alt} 
\ee
where $\hcoeff$ is the function from \eq\nr{cal_F}.
In addition, we may vary~$t^{ }_1$, 
for instance in the range 
$k/[a^{ }_1 H(t^{ }_1)] \in 10^2_{ } ... 10^3_{ }$.\footnote{%
 The value has to be large enough so that we are safely within
 the Hubble horizon at $t=t^{ }_1$, but not so large that oscillations
 would be prohibitively difficult to handle numerically. 
 }
We find that in this range, 
the difference between \eqs\nr{initial_dot_R_varphi}
and \nr{initial_dot_R_varphi_alt} has only
a minor influence on the solution, and that approaching
$k/[a^{ }_1 H(t^{ }_1)] \sim 10^3_{ }$, the dependence
on $t^{ }_1$ becomes insignificant, provided that
$\Upsilon(t^{ }_1) \ll H(t^{ }_1)$.

%
\subsection{Handling singularities}
\la{ss:singularities}

%
\paragraph{Identification of a problem.}

We note from \eq\nr{cal_F} that 
when  $\dot{\bar\varphi}$ crosses zero, 
the coefficient~$\hcoeff$
becomes singular.\footnote{%
 Two other singular terms in \eq\nr{dot_R_varphi}
 are the noise $-\varrho H/\dot{\bar\varphi}$,
 to which we return in \se\ref{ss:noise}, and the coefficient 
 $V^{ }_{\der\varphi\T}/(\dot{\bar\varphi}\,\bar{e}^{ }_{\der\T})$, 
 which can be regularized in the same way as $\mathcal{F}$.
 }
Specifically, let us inspect a negative minimum of~$\bar\varphi$,
around an ``oscillation'' time $t^{ }_o$, 
\be
 \bar\varphi = - A \cos[m(t-t^{ }_o)]
 \;,\quad 
 \dot{\bar\varphi} = A\, m \sin[m(t-t^{ }_o)]
 \;. \la{osc_soln}
\ee
Then, according to \eq\nr{cal_F},   
\be
 \hcoeff  
 \;\supset\;
 \frac{\ddot{\bar\varphi}}{\dot{\bar\varphi}}
 \;\stackrel{\; t\,\approx\, t^{ }_o}{\approx}\;
  \frac{1}{t-t^{ }_o}
 \;. \la{pole}
\ee

In order to deal with the singularity, we may regularize it. 
Assume first that we replace $1/\dot{\bar\varphi}$ by
a principal value, 
$
 \dot{\bar\varphi}/(\dot{\bar\varphi}^2_{ } + \delta^2_{ })
$.
Then
$
 \hcoeff \approx (t-t^{ }_o) / [(t-t^{ }_o)^2_{ } + \hat\delta^2_{ }]
$, 
with $\hat{\delta} \equiv \delta / (A\, m^2_{ })$, and the system is 
integrable. The question is, how does it behave when $\hat\delta\to 0$?

The most problematic term comes from the first line of \eq\nr{dot_R_varphi}, 
\be
 \ddot{\mathcal{R}}^{ }_\varphi
 \approx 
 - \frac{2 (t-t^{ }_o)}{ (t-t^{ }_o)^2_{ } + \hat\delta^2_{ } } 
 \,\dot{\mathcal{R}}^{ }_\varphi
 \;. \la{problem}
\ee
As a homogeneous second-order differential equation, this has two 
independent solutions. There is the trivial one, 
${\mathcal{R}}^{ }_\varphi = \mbox{constant}$, and this in fact represents
the physical behaviour of the system. But there is also a second
solution, 
\ba
 \dot{\mathcal{R}}^{ }_\varphi & = & 
 \frac{c}{(t-t^{ }_o)^2_{ } + \hat\delta^2_{ }}
 \;, \la{prblm_1} \\[2mm]
 \mathcal{R}^{ }_\varphi(t^{ }_+) - \mathcal{R}^{ }_\varphi(t^{ }_-)
 & = & 
 \frac{c}{\hat\delta}
 \biggl[\, 
 \arctan\biggl( \frac{t^{ }_+ - t^{ }_o}{\hat\delta} \biggr) - 
 \arctan\biggl( \frac{t^{ }_- - t^{ }_o}{\hat\delta} \biggr)
 \,\biggr]
 \;\; 
 \overset{ t^{ }_+ - t^{ }_o \, \gg \, +\hat\delta }{  
  \underset{ t^{ }_- - t^{ }_o \, \ll \, -\hat\delta }{ \approx } } 
 \;\; 
 \frac{c\pi}{\hat\delta}
 \;. \hspace*{6mm} \la{prblm_2}
\ea
In other words, the second solution of 
$\mathcal{R}^{ }_\varphi$ undergoes a large jump when
$\dot{\bar\varphi}$ crosses zero. 
For this jump, the limit $\hat\delta\to 0$ does not exist. 

The concrete problem is that even though this second 
solution should not be realized (i.e.\ $c=0$), in a practical numerical 
integration it will sooner or later appear, due to accumulating numerical 
errors. Once this happens, there is a huge jump the next time that
$\dot{\bar\varphi}$ crosses zero, and 
afterwards the numerical solution is meaningless. 
 
%
\paragraph{Solution: complexified variables.}

A simple solution to the instability is offered by the 
observation that due to the matching to quantum-mechanical initial
conditions (cf.\ \se\ref{ss:initial}), 
$\mathcal{R}^{ }_\varphi$ is a complex variable. Taken
literally, \eqs\nr{dot_R_varphi}--\nr{dot_S_T} do not mix the real
and imaginary parts of $\mathcal{R}^{ }_\varphi$. 
However, having available both parts 
can be turned into an advantage, 
when we regularize $1/\dot{\bar\varphi}$.

Specifically, suppose that we displace 
the pole in $\dot{\bar\varphi}$ off the real axis, 
\be
 \frac{1}{\dot{\bar\varphi}} \; \longrightarrow \; 
 \frac{1}{\dot{\bar\varphi} + i \delta}
 \;, \quad \delta \in \mathbbm{R}
 \;. \la{regularization}
\ee
Then \eq\nr{problem} gets replaced with 
\be
 \ddot{\mathcal{R}}^{ }_\varphi
 \approx 
 - \frac{2}{ t-t^{ }_o + i \hat\delta } 
 \,\dot{\mathcal{R}}^{ }_\varphi
 \;, \la{no_problem}
\ee
which leads to the second solution 
\ba
 \dot{\mathcal{R}}^{ }_\varphi & = & 
 \frac{c}{(t-t^{ }_o  + i \hat\delta)^2_{ } }
 \;, \\[2mm]
 \mathcal{R}^{ }_\varphi(t^{ }_+) - \mathcal{R}^{ }_\varphi(t^{ }_-)
 & = & 
 c\,
 \biggl[\, 
    \frac{1}{ t^{ }_- - t^{ }_o + i \hat\delta} - 
    \frac{1}{ t^{ }_+ - t^{ }_o + i \hat\delta} 
 \,\biggr]
 \;\; 
 \overset{ t^{ }_+ - t^{ }_o \, \gg \, +\hat\delta }{  
  \underset{ t^{ }_- - t^{ }_o \, \ll \, -\hat\delta }{ \approx } } 
 \;\; 
 0 
 \;. \hspace*{6mm} \la{no_prblm_2}
\ea

Equation~\nr{no_prblm_2} implies that even if we excited the second
solution through numerical errors (i.e.\ $c\neq 0$), 
the uncertainty does not get amplified
when we cross singularities. Therefore, the numerical solution is stable. 
In practice, we have set $\delta \simeq 10^{-15}_{ } \mpl^2$, and
checked that our results are independent of this choice.

%
%

%
\subsection{Averaging over fast background oscillations}
\la{ss:matter}

Let us inspect the background evolution at late times. 
Inserting the pressure $\bar p$ from \eq\nr{p_mixed} and
the energy density $\bar e$ from \eq\nr{e_mixed}, 
\eqs\nr{bg_t}
and \nr{bg_Tmunu_new} can be expressed as  
\ba
 \ddot{\bar\varphi} + 
 \bigl( 3 H + \Upsilon \bigr)\,\dot{\bar\varphi} +  V^{ }_{\der\varphi} 
 & = & 
  0
 \;, \la{eom_field} \\[3mm] 
 \dot{e}^{ }_r + 3 H \bigl( e^{ }_r + p^{ }_r - T  V^{ }_{\der\T} \bigr)
 - T \dot{V}^{ }_{\der\T}  
 & = &
 \Upsilon^{ }
 \dot{\bar\varphi}^2
 \;, \la{eom_plasma} \\[1mm]
 H 
 & = & 
 \sqrt{\frac{8\pi}{3}} 
 \frac{\sqrt{ 
 \tfr{1}{2}\! \dot{\bar\varphi}^2 
  +
  e^{ }_r + V -T  V^{ }_{\der\T}  
  }}
      { \mpl^{ } }
 \;. \la{eom_friedmann} 
\ea

Now, as time goes by, the inflaton settles around 
the minimum of the potential. Let us assume that the minimum
is at $\bar\varphi = 0$. Then, \eq\nr{eom_field} becomes the
oscillator equation of motion, 
$
 \ddot{\bar\varphi} + 
 \bigl( 3 H + \Upsilon \bigr)\,\dot{\bar\varphi} +  
 m^2_{ }\bar\varphi 
  \approx
  0
$, 
where $m^2_{ } \equiv V^{ }_{\der\varphi\varphi} |^{ }_{\bar\varphi = 0}$.
Once $3 H + \Upsilon \ll m$, this implies the presence of rapid
oscillations. It is numerically expensive to integrate over
such oscillations for a long period of time, 
and a more efficient method is needed. 

We note in passing that rapid oscillations require some care
as far as the validity of the formalism is concerned. The plasma 
has its own equilibration rate, let us call it $\Gamma$. 
If $\Gamma \gg m$, the plasma stays in thermal equilibrium as the
inflaton evolves. If $\Gamma \ll m$, the plasma does not have time
to react to every change of the inflaton field. Instead, it feels
the average value of $\bar\varphi$-dependent quantities, 
and continues to stay equilibrated. The delicate
case is when $\Gamma \sim m$, because then the plasma would like to
adjust to changes in the inflaton field, but does not quite have 
time to do so, and the system falls out of equilibrium (via~hysteresis).
However, this last case takes place
only in a narrow temperature range, so in practice
we assume the plasma to be thermal all the time. 

In order to obtain equations for the oscillation regime, 
we make the simplifying assumption that the potential is 
temperature-independent, $V^{ }_{\der\T} \simeq 0$
(cf.\ footnote~\ref{fn:warm0} on p.~\pageref{fn:warm0}). 
Let us define the inflaton background energy density as 
\be
 e^{ }_{\bar\varphi} \; \stackrel{V^{ }_{\der\T}\,\simeq\, 0}{\equiv} \; 
 \frac{\dot{\bar\varphi}^2_{ }}{2} + V
 \;. \la{ephi_def}
\ee
Multiplying \eq\nr{eom_field} with $\dot{\bar\varphi}$, we then find
that $e^{ }_{\bar\varphi}$ is approximately conserved, 
up to the loss term
$
 -(3H + \Upsilon) \dot{\bar\varphi}^2_{ }
$.
However, which of the terms in \eq\nr{ephi_def} 
carries the energy density, 
changes rapidly. On average, both terms contribute equally, 
\be
 \biggl\langle\,
 \frac{\dot{\bar\varphi}^2_{ }}{2} 
 \,\biggr\rangle^{ }_{\Delta t\, \sim\, \frac{2\pi} m}
 \;\approx\;
 \bigl\langle\,
 V
 \,\bigr\rangle^{ }_{\Delta t\, \sim\, \frac{2\pi} m}
 \quad \Rightarrow \quad 
 e^{ }_{\bar\varphi} 
 \; \approx \; 
 \bigl\langle\,
 \dot{\bar\varphi}^2_{ } 
 \,\bigr\rangle^{ }_{\Delta t\, \sim\, \frac{2\pi} m} 
 \;. \la{aves}
\ee
Averaging the background equations over $\Delta t$, 
they can be turned into
\ba
 \dot{e}^{ }_{\bar\varphi} + 3 H\, e^{ }_{\bar\varphi}
 & \simeq & -\,\Upsilon \, e^{ }_{\bar\varphi}
 \;, \la{eom_ephi} \\[2mm]
 \dot{e}^{ }_r + 3 H\,\bigl(\, e^{ }_r + p^{ }_r \,\bigr)
 & \simeq & +\,\Upsilon\, e^{ }_{\bar\varphi}
 \;, \la{eom_er} \\[2mm]
 H 
 & \simeq &
 \sqrt{\frac{8\pi}{3}} 
 \frac{\sqrt{ 
  e^{ }_{\bar\varphi}
  +
  e^{ }_r 
  }}
      { \mpl^{ } }
 \;. \la{eom_hubble} 
\ea
If $e^{ }_{\bar\varphi} \gg e^{ }_r$, 
the solution is referred to as a matter-dominated epoch 
(cf.,\ e.g., ref.~\cite{matter}).

In practice, while solving \eqs\nr{eom_field}--\nr{eom_friedmann}, 
we count how many times $\bar\varphi$ changes its sign. 
After about 20 sign flips, 
we fix the initial condition for $e^{ }_{\bar\varphi}$ 
from \eq\nr{ephi_def}, and then move on to solve 
\eqs\nr{eom_ephi}--\nr{eom_hubble}. The accuracy of the procedure
can be checked by changing the number of sign flips, 
as well as the switching criterion, e.g.\ whether
from zeros of $\bar{\varphi}$ or $\ddot{\bar{\varphi}}$
(we do not use the zeros of~$\dot{\bar{\varphi}}$, because the
singularity of \se\ref{ss:singularities} momentarily kicks the 
solution off its proper trajectory). We find the 
residual errors to normally be on the percent level. 

\vspace*{3mm}

Turning to the equations for perturbations in
\eqs\nr{dot_R_varphi}--\nr{dot_S_T},
they can also be converted into the regime
of \eqs\nr{ephi_def}--\nr{eom_hubble}. Recalling the assumption
of a temperature-independent potential 
$
 V \simeq \frac{1}{2} m^2_{ } \varphi^2_{ }
$
near $\bar\varphi = 0$, and employing \eq\nr{aves}, 
we can just replace 
\be
 \dot{\bar\varphi}^2_{ }\to e^{ }_{\bar\varphi}
 \;, \quad
 \bar{e} \to e^{ }_r + \frac{e^{ }_{\bar\varphi}}{2}
 \;, \quad
 \bar{p} \to p^{ }_r - \frac{e^{ }_{\bar\varphi}}{2}
 \;, \quad
 \bar{e}^{ }_{\der\T} \to e^{ }_{r\der\T}
 \;, \quad
 \bar{p}^{ }_{\der\T} \to p^{ }_{r\der\T}
 \;, \quad
 V^{ }_{\der\varphi\T} \to 0
 \;.  \la{fast_oscs}
\ee
These fix most of the coefficients, 
apart from $\hcoeff$ from \eq\nr{cal_F} and 
$\dot{\bar{e}}\supset \bar{e}^{ }_{\der\varphi}\,\dot{\bar\varphi}$
from \eq\nr{Delta_v}, which oscillate;
and the noise terms in \eqs\nr{dot_R_varphi}
and \nr{dot_S_T}, to which we return
in \se\ref{ss:noise}.

The oscillating part of $\hcoeff$ reads, with the regularization
from \eq\nr{regularization}, 
\be
 \hcoeff \supset 
 \frac{\ddot{\bar\varphi}}{\dot{\bar\varphi} + i \delta}
 \;. 
\ee
An average over an oscillation period, in the sense of
\eq\nr{aves}, gives 
\be
 \biggl\langle\, 
   \frac{\ddot{\bar\varphi}}{\dot{\bar\varphi} + i\delta}
 \,\biggr\rangle^{ }_{\Delta t\, \sim\, \frac{2\pi} m}
 = 
 \frac{m}{2\pi}
 \int_0^{\frac{2\pi}{m}} \! {\rm d}t\, 
 \frac{\ddot{\bar\varphi}}{\dot{\bar\varphi} + i\delta}
 = 
 \frac{m}{2\pi}
 \biggl[\,
  \ln\bigl( \dot{\bar\varphi} + i \delta \bigr)  
 \,\biggr]^{ \frac{2\pi}{m} }_{0}
 = 
 0
 \;. \la{osc_ave_1}
\ee
Therefore, we obtain a slowly evolving expression
also for $\hcoeff$, 
\be
 \hcoeff \;\to \; 
 \frac{4\pi G (\, e^{ }_{\bar\varphi} + {e}^{ }_r + {p}^{ }_r \,) }{H}
 \;. \la{cal_F_2nd} 
\ee
As for the evolution of the energy density via changes of 
$e^{ }_{\bar\varphi}$, we find
\be
 \bar{e}^{ }_{\der\varphi}\,\dot{\bar\varphi}
 \; \stackrel{\rmii{\nr{fast_oscs}}}{\to}   \;
     \frac{\dot{e}^{ }_{\bar\varphi}}{2}
 \; \stackrel{\rmii{\nr{eom_ephi}}}{\simeq} \; 
   - \frac{(3H+\Upsilon)e^{ }_{\bar\varphi}}{2}
 \;.
\ee

%
\subsection{Going over to a radiation-dominated universe}
\la{ss:radiation}

At a late stage of the evolution, when $\Upsilon \gg H$, the background
energy density $e^{ }_{\bar\varphi}$ starts to decrease exponentially. 
Soon this leads to $e^{ }_{\bar\varphi} \ll e^{ }_r$, and we enter
a radiation-dominated epoch. It may be questioned what the physical
meaning of $\mathcal{R}^{ }_\varphi$ is, when the corresponding 
background quantity $e^{ }_{\bar\varphi}$ has ceased to play a role. 
Here we show that $\mathcal{R}^{ }_\varphi$ 
decouples from the evolution of 
$\mathcal{R}^{ }_v$ and $\mathcal{R}^{ }_\T$
in this regime. 

Let us first derive the equations for the case
that no $\varphi$ is present in the first place.
Proceeding like in \se\ref{sss:combined_R}, but omitting 
$\bar\varphi$ and $\delta\varphi$, the starting point is given
by the last few terms in \eqs\nr{dTmunu_i} and \nr{dTmunu_0}. 
We can again eliminate 
$h^{ }_0$, $h_0'$ and 
$h_0' + 3 h_\rmii{D}' + \nabla^2 h$
by employing 
\eqs\nr{pert_einstein_0i_new} and \nr{from_einstein_sum_new}.
The variables $h-v$ and $\delta T$ can be expressed in terms
of $\mathcal{R}^{ }_v$ and $\mathcal{R}^{ }_\T$ via 
\eqs\nr{def_R_v} and \nr{def_R_T}, respectively. 
An important crosscheck is that
subsequently the many 
appearances of $h^{ }_\rmiii{D} + \frac{\nabla^2\vartheta}{3}$
cancel exactly. Making use of the background identities from 
\eqs\nr{bg_t} and \nr{bg_Tmunu_new} as well as
thermodynamic relations from \se\ref{sss:Tmunu}, we obtain
\ba
 \dot{\mathcal{R}}^{ }_v & = & 
 \underbrace{ 
 \frac{\dot{\bar{p}}}{\bar{e}+\bar{p}}
 }_{ 
  \dot{T} / T
 }
 \,\bigl(\, \mathcal{R}^{ }_\T - \mathcal{R}^{ }_v \,\bigr)
 \;, \la{dot_R_v_rad} \\
 \dot{\mathcal{R}}^{ }_\T & = & 
 \frac{\mathcal{R}^{ }_v}{3 H} \frac{k^2_{ }}{a^2_{ }} 
 \;
 \underbrace{ 
 - \;
 \frac{4\pi G(\bar{e} + \bar{p})   
 }{H}
 }_{ 
  \dot{H} / H
 }
 \,\bigl(\, \mathcal{R}^{ }_\T - \mathcal{R}^{ }_v \,\bigr)
 \;. \la{dot_R_T_rad} 
\ea 
These equations manifest the feature discussed below
\eq\nr{dot_S_T}, namely that if $k/a \ll H$, the system has 
a stationary solution $\mathcal{R}^{ }_v = \mathcal{R}^{ }_\T$.
Furthermore, since the equations are of first order in time, 
it is sufficient to know the common value of these variables
at an initial moment. 
We also note that in terms of 
$\mathcal{R}^{ }_v$ and $\mathcal{R}^{ }_\T$
and in the absence of $e^{ }_{\bar\varphi}$, 
the energy density perturbation
from \eq\nr{Delta_v} becomes
\be
 \Delta^{ }_v
 \; \approx \;  
 \frac{\dot{\bar e}}{ \bar e H }
 \bigl(\, \mathcal{R}^{ }_v - \mathcal{R}^{ }_\T \,\bigr)
 \la{Delta_v_new}
 \;.
\ee
In practice 
we have gone over to \eqs\nr{dot_R_v_rad}--\nr{Delta_v_new} when
$e^{ }_{\bar\varphi} < e^{ }_r/100$.

We now show how 
\eqs\nr{dot_R_v_rad} and \nr{dot_R_T_rad} arise as a limit 
of \eqs\nr{dot_R_varphi}--\nr{dot_S_T}.
When $\dot{\bar\varphi}^2_{ }\ll \bar{e}$, the background 
identities from \eqs\nr{bg_t} and \nr{bg_Tmunu_new}
read 
$
 \dot{H} = -4\pi G (\bar{e}+\bar{p})
$, 
$
 3 H^2_{ } = 8\pi G \bar{e}
$, 
$
 \dot{\bar{e}} = - 3 H (\bar{e} + \bar{p}) 
$.
As there is only one dynamical background quantity ($T$), we can write
$
 \E^{ }_\T = \bar{e}^{ }_{\der\T} \dot{T}\,
 (\mathcal{R}^{ }_\T - \mathcal{R}^{ }_\varphi)
 = \dot{\bar{e}}\,
 (\mathcal{R}^{ }_\T - \mathcal{R}^{ }_\varphi)
 = 
  - 3 H (\bar{e} + \bar{p})
 (\mathcal{R}^{ }_\T - \mathcal{R}^{ }_\varphi)
$, 
as well as 
$
 \bar{p}^{ }_{\der\T} / \bar{e}^{ }_{\der\T} = \dot{\bar{p}}/\dot{\bar{e}}
$.
Inserting these, the coefficients of 
$\dot{\mathcal{R}}^{ }_\varphi$ and 
$\mathcal{R}^{ }_\varphi$ vanish in  
\eqs\nr{dot_S_v} and \nr{dot_S_T}, and 
\eqs\nr{dot_R_v_rad} and \nr{dot_R_T_rad} are reproduced.
In other words, the evolution of $\mathcal{R}^{ }_\varphi$ decouples
from those of $\mathcal{R}^{ }_v$ and $\mathcal{R}^{ }_\T$, and 
$\mathcal{R}^{ }_\varphi$ 
no longer affects physical quantities, like 
$\Delta^{ }_v$ in \eq\nr{Delta_v_new}. 

%
\subsection{What about the noise terms?}
\la{ss:noise}

Even though we have mostly not considered the effects induced by
the noise terms in the present study, let us nevertheless verify that
they are non-singular in the sense of \se\ref{ss:singularities}, and 
have a well-defined limit when going to the 
regimes of \ses\ref{ss:matter} and \ref{ss:radiation}.

According to \eq\nr{varphi_noise}, the noise appearing
in \eqs\nr{dot_R_varphi} and \nr{dot_S_T} has the autocorrelator
\be
 \bigl\langle\, 
   \varrho^{ }(t^{ }_1,\vec{k}) \, 
   \varrho^{ }(t^{ }_2,\vec{q})
 \,\bigr\rangle
 \;
    =  
 \; 
 \, \delta(t^{ }_1 - t^{ }_2)
 \; \deltabar(\vec{k+q})  
 \, 
 \frac{ \Omega 
       }{a^3_{ }}
 \; + 
 \; \rmO(\delta^3_{ })
 \;. \la{noise_physical}
\ee
After the regularization
from \eq\nr{regularization}, 
and noting that \eq\nr{noise_physical} implies
that the functions appearing in the autocorrelator are evaluated at the 
same time, we see that the autocorrelator of the noise affecting
${\mathcal{R}}^{ }_\varphi$ 
is proportional to  
$
 \Omega H^2_{ } / (\dot{\bar\varphi} + i \delta)^2_{ }
$, 
and requires a careful discussion (see below).
In contrast, 
that affecting ${\E}^{ }_\T$
is proportional to  
$
 \Omega\, \dot{\bar\varphi}^2_{ }H^2_{ }
$, and requires no regularization.
The value for the regime of 
fast scalar oscillations is obtained directly from 
\eq\nr{fast_oscs}, via the replacement 
$\dot{\bar\varphi}^2 \to e^{ }_{\bar\varphi}$.

For specifying the noise for
${\mathcal{R}}^{ }_\varphi$, 
we need to consider separately two regimes: 
\bi

\item[(i)] 
At early times, when our background variable is $\bar\varphi$, the noise
is complex. As our field space for perturbations has been complexified, 
there is nothing special about this. 

\item[(ii)] 

In the regime of fast oscillations, corresponding to 
\se\ref{ss:matter}, we need to evaluate an average like 
in \eq\nr{osc_ave_1}. This can be done with contour integrals.
Parametrizing the solution like in \eq\nr{osc_soln},  
we write 
$
 \sin[m(t-t^{ }_o)] = (z - z^{-1}_{ })/(2 i)
$
with 
$
 z = e^{im(t-t^{ }_o)}_{ }
$,
and take then $z$ as an integration variable. 
This yields
\be
 \biggl\langle \frac{1}{\dot{\bar\varphi} + i\delta}
 \biggr\rangle^{ }_{\Delta t\, \sim\, \frac{2\pi} m}
 \; = \; -\, 
 \frac{i \sign(\delta)} 
  {\bigl( A^2_{ } m^2_{ } + \delta^2_{ } \bigr)^{1/2}_{ }}
 \;. 
\ee
Taking a derivative with respect to $\delta$ subsequently produces
\be
 \biggl\langle \frac{1}{(\dot{\bar\varphi} + i\delta)^2_{ }}
 \biggr\rangle^{ }_{\Delta t\, \sim\, \frac{2\pi} m}
 \; \stackrel{\delta\,\neq\,0}{=} \; -\,
 \frac{|\delta|}{ \bigl( A^2_{ }m^2_{ } + \delta^2_{ } \bigr)^{3/2}_{ } }
 \;. 
\ee
If we send $\delta\to 0^+_{ }$, which is possible in the 
regime of fast scalar oscillations, because all singular coefficients
have disappeared, we see that the noise autocorrelator vanishes. 
Therefore, the noise can be omitted from the evolution 
of~$\mathcal{R}^{ }_\varphi$ in this regime. 

\ei

\noindent
To summarize, in the regime of fast scalar oscillations, 
noise can be eliminated from~$\mathcal{R}^{ }_\varphi$, but it 
does remain present in the evolution of $\E^{ }_\T$, as
long as $e^{ }_{\bar\varphi} \neq 0$. 
In the regime of \se\ref{ss:radiation}, 
where $\varphi$ has completely
disappeared, no noise remains present within our setup.

%
\section{Numerical illustrations}
\la{se:numerics}

We now turn to numerical benchmarks with which we illustrate the
solution of our coupled set of equations for the scalar perturbations.

%
\subsection{Background solutions}
\la{se:num_background}

We start by defining a concrete model 
for which we determine the background solution. 
This is inspired by a non-Abelian axion inflation framework, 
argued to possess a thermal fixed point~\cite{fixed_pt_1,fixed_pt_2}, 
however we simplify the 
friction coefficient $\Upsilon$ and the radiation thermodynamic 
functions $e^{ }_r$ and  $p^{ }_r$~\cite{eos,Tmax}, in order to obtain
a scenario which is easy to reproduce and 
has its relevant physics taking place
within a short time interval. 

For the inflaton potential, we adopt an instanton-induced ansatz, 
\be
 V
 \;\equiv\;
 m^2 f_a^2\, 
 \biggl[ 1 - \cos\biggl( \frac{\bar\varphi}{f^{ }_a} \biggr) \biggr]
 \;. \la{V}
\ee
The parameters are fixed in accordance with Planck data for CMB
observables 
(specifically $A^{ }_s$, $n^{ }_s$, $r$ at the $2\sigma$ level), as 
\be
 f^{ }_a = 1.25 \, \mpl^{ } 
 \;, \quad 
 m = 1.09 \times 10^{-6}_{ }\, \mpl^{ }  
 \;, \la{params}
\ee
whereby constrains from high-frequency gravitational wave production 
(LISA projections and $N^{ }_\rmi{eff}$) 
are respected as well~\cite{scan}.
Following the arguments in ref.~\cite{Tmax}, we treat these parameters,
and therefore $V$, as temperature independent, i.e.\ $V^{ }_{\der\T} = 0$.

In the setup assumed, 
$\sqrt{ m f^{ }_a} \sim \Lambda^{ }_\rmii{UV} 
\sim 10^{-3}_{ } \mpl^{ }$
corresponds to the confinement scale of some unified theory. 
We assume inflationary physics to take place at scales much 
below~$\Lambda^{ }_\rmii{UV}$, so that the UV theory is confined and 
its excitations are heavy. 
At the same time, to have a concrete ansatz for the radiation bath, 
the inflaton is assumed to also 
couple to an unbroken SU($\Nc^{ }$) gauge theory, 
with a confinement
scale $\Lambda^{ }_\rmii{IR} \ll \Lambda^{ }_\rmii{UV}$. 
We assume inflationary physics to take 
place at scales much higher than $\Lambda^{ }_\rmii{IR}$, so that the 
second non-Abelian radiation component is weakly coupled.
The assumption $\Lambda^{ }_\rmii{IR} \ll \Lambda^{ }_\rmii{UV}$ 
justifies the potential in \eq\nr{V} 
and leads to the qualitative form of the friction 
that will be introduced in \eq\nr{Upsilon_ansatz}.    

To sustain an extended period of inflation, the initial value of the 
field is chosen close to the maximum of the potential, for instance
\be
 \bar\varphi(t^{ }_\rmii{ref}) = 3.5 \,\mpl^{ }
 \;, \quad
  t^{ }_\rmi{ref} 
  \; \equiv \; 
  \sqrt{\frac{3}{4\pi}} 
  \frac{ m^{ }_\rmiii{pl} }{m\, \bar\varphi(t^{ }_\rmiii{ref})}
 \;. \la{t_ref} 
\ee
If the potential were quadratic, $t^{ }_\rmi{ref}$ would
correspond to the initial inverse Hubble rate; as it is not, 
this holds only in order of magnitude, 
$t^{ }_\rmi{ref} \sim H^{-1}_{ }(t^{ }_\rmi{ref})$. 
Given the long time period that needs to be 
considered, it is helpful in practice
to replace~$t$ as an integration variable through
$
 x \;\equiv\; 
 \ln\bigl( { t } / { t^{ }_\rmi{ref} } \bigr)
$.

For the radiation, 
we adopt the values corresponding to a weakly coupled SU(3) 
Yang-Mills plasma, 
\be
 e^{ }_r 
 \; = \; 
 \frac{ g^{ }_* \pi^2 T^4 }{ 30 }
 \;, \quad
 p^{ }_r 
 \; = \; 
 \frac{ g^{ }_* \pi^2 T^4 }{ 90 }
 \;, \quad
 g^{ }_* = 16 
 \;. 
\ee
The initial temperature is taken to be small, 
$T(t^{ }_\rmi{ref}) \ll H(t^{ }_\rmi{ref})$, but it rapidly
increases and settles to an almost stationary fixed point. 

The final key ingredient for \eqs\nr{eom_field} and \nr{eom_plasma} 
is the friction coefficient, $\Upsilon$. For the case of a pseudoscalar
field interacting with a Yang-Mills plasma, this can be 
estimated numerically~\cite{shape}.
However, for purposes of illustration, it is helpful to rather
consider a model, having a similar overall shape as 
suggested by the numerical data, 
but a larger magnitude, 
so that the physics of equilibration takes 
place faster. We do this by introducing 
\be
 \Upsilon 
 \; \equiv \; 
 \frac{
    \kappa^{ }_\T\, (\pi T)^3_{ } + 
    \kappa^{ }_m\, m^3_{ }
 }{(4\pi)^3_{ } f_a^2}
 \;,  \la{Upsilon_ansatz}
\ee
where the coefficients $\kappa^{ }_\T$ and $\kappa^{ }_m$ will
be set to quite large values. 
Concretely, we consider three cases, defined as
\ba
 \mbox{(a)}: && 
 \kappa^{ }_\T \; = \; 10^6_{ }
 \;, \quad
 \kappa^{ }_m \; = \;  10^8_{ }
 \;, \la{case_v_a} \\
 \mbox{(b)}: && 
 \kappa^{ }_\T \; = \; 10^7_{ }
 \;, \quad
 \kappa^{ }_m \; = \;  10^9_{ }
 \;, \la{case_v_b} \\
 \mbox{(c)}: && 
 \kappa^{ }_\T \; = \; 10^6_{ } 
 \;, \quad
 \kappa^{ }_m \; = \;  10^{10}_{ }
 \;. \la{case_v_c} 
\ea
The corresponding results are shown in \fig\ref{fig:cases}.\footnote{%
 It should be stressed that 
 from the point of view of the microscopic theory, 
 the couplings in \eqs\nr{case_v_a}--\nr{case_v_c} 
 are unnaturally large. 
 A realistic system could go in this direction  
 if a Yang-Mills theory 
 is strongly coupled at the energy scale of inflation~\cite{Tmax}, 
 however then it is difficult to 
 determine $\Upsilon$ and $V$ quantitatively. 
 We have also studied couplings more appropriate
 for a weakly-coupled Yang-Mills plasma,
 $\kappa^{ }_\T \sim \kappa^{ }_m \sim \rmO(1)$,
 confirming that the features remain the same, 
 however then the matter domination period lasts much longer. 
 } 
Despite the large coefficients, all solutions
remain in the weak regime, i.e.\ with $\Upsilon \ll H$ during
the inflationary period (this observation is consistent
with the recent analysis in ref.~\cite{szell}). 

\begin{figure}[t]

\vspace*{-0.1cm}

\centerline{%
   \epsfysize=5.0cm\epsfbox{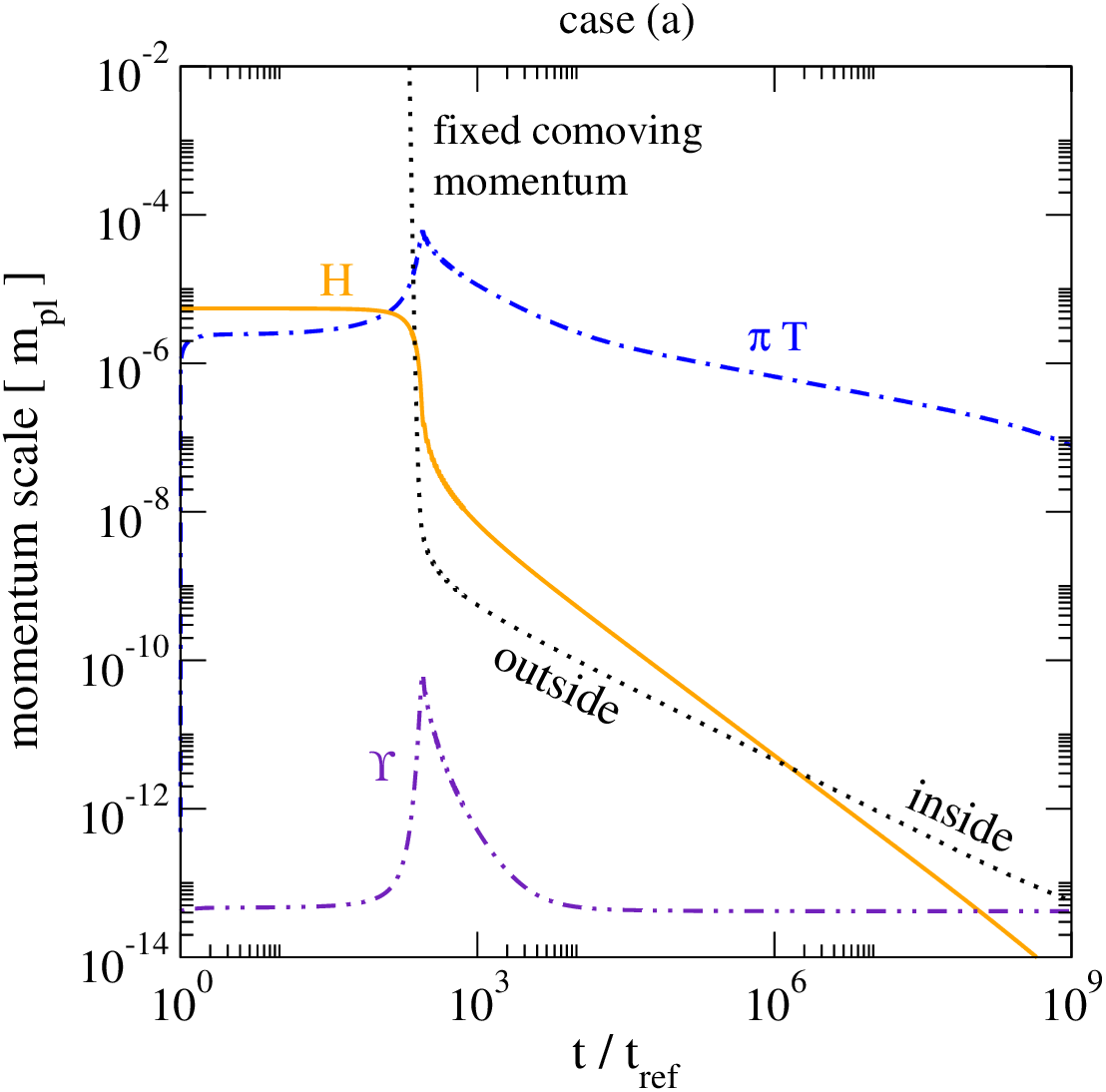}
   ~~~\epsfysize=5.0cm\epsfbox{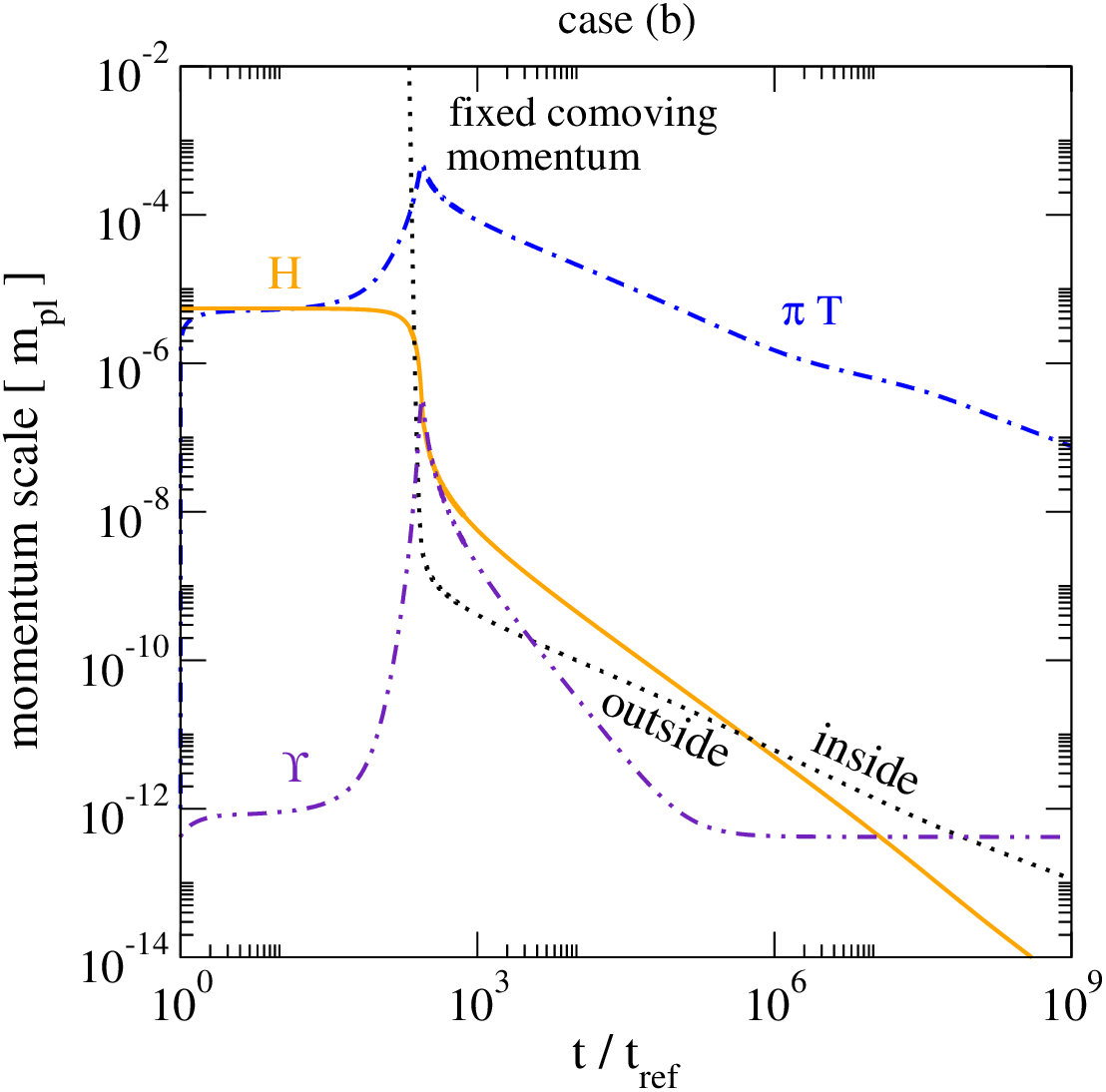}
   ~~~\epsfysize=5.0cm\epsfbox{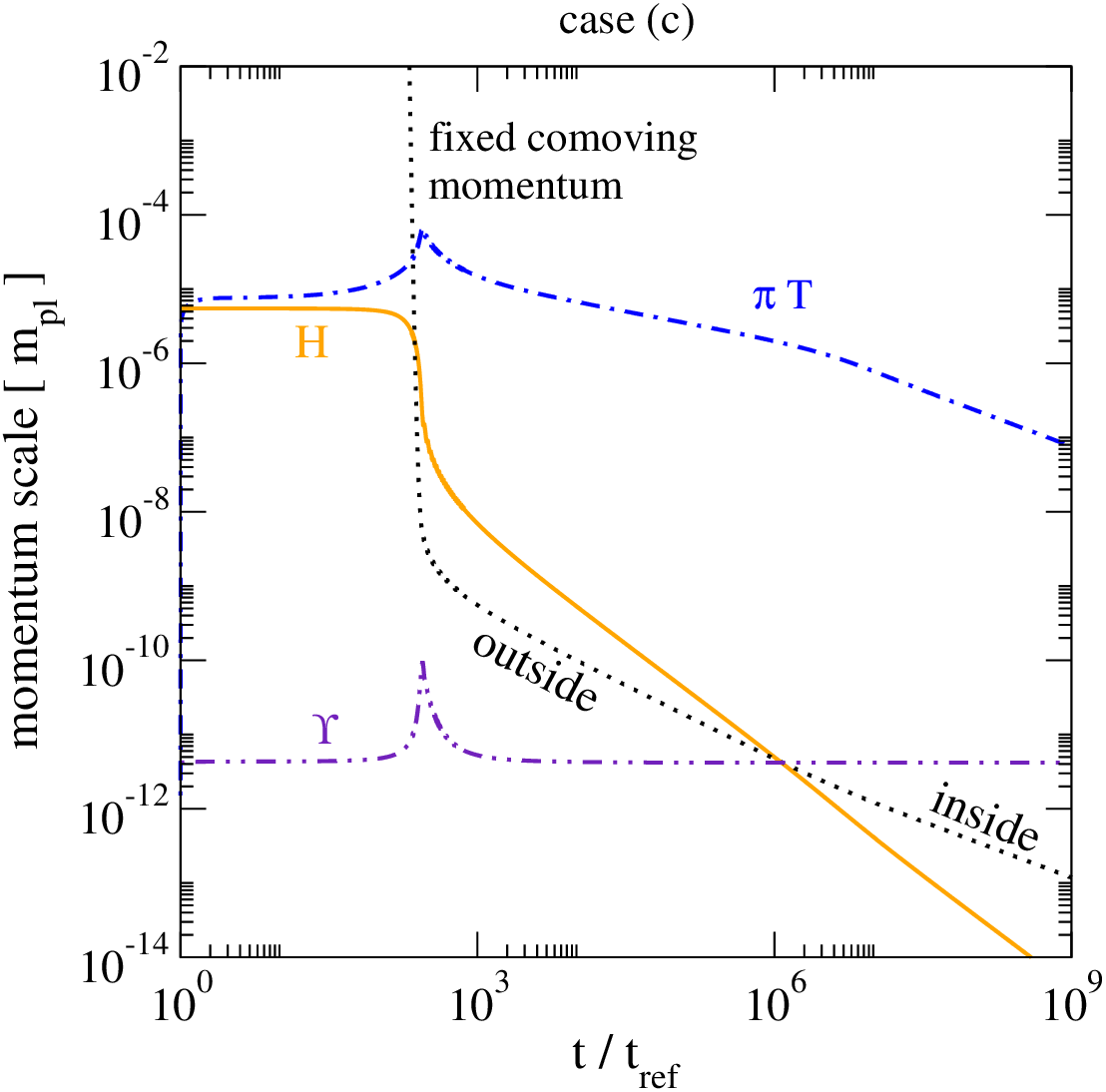}
}

\caption[a]{\small
  Examples of background solutions, with couplings
  fixed according to \eqs\nr{case_v_a}--\nr{case_v_c}.
  The comoving momentum is illustrated with a mode 
  which takes the value $10^{-10}_{ }\mpl$
  at $t/t_\rmii{ref}^{ } = 10^4_{ }$. 
}

\la{fig:cases}
\end{figure}

In case~(a), with $\Upsilon < H$, the inflaton keeps on oscillating
for a long time, leading to a prolonged
matter-dominated epoch. 
In case~(b), the
thermal part of $\Upsilon$ is large, 
leading to a high maximal temperature. In this case, 
the expansion shows a radiation-dominated epoch soon after inflation.
Finally, 
in case~(c), the
vacuum contribution to $\Upsilon$, 
which can be identified as the time-independent
(horizontal) part of $\Upsilon$ in \fig\ref{fig:cases}, exceeds the Hubble
rate at $t \simeq 10^6_{ }\, t^{ }_\rmi{ref}$. 
After this moment, the inflaton energy
density $e^{ }_{\bar\varphi}$ decreases rapidly, and $e^{ }_r$ becomes the
dominant energy component. Therefore we enter a radiation-dominated epoch, 
and the temperature starts to decrease faster. 

%
\subsection{Curvature and energy density perturbations}
\la{ss:num_R}

We now turn to the solution of \eqs\nr{dot_R_varphi}--\nr{dot_S_T}.
The noise $\varrho$ will be omitted, whereby 
we are considering only dissipative phenomena.

\begin{figure}[t]

\vspace*{-0.1cm}

\centerline{%
   ~~~\epsfysize=5.0cm\epsfbox{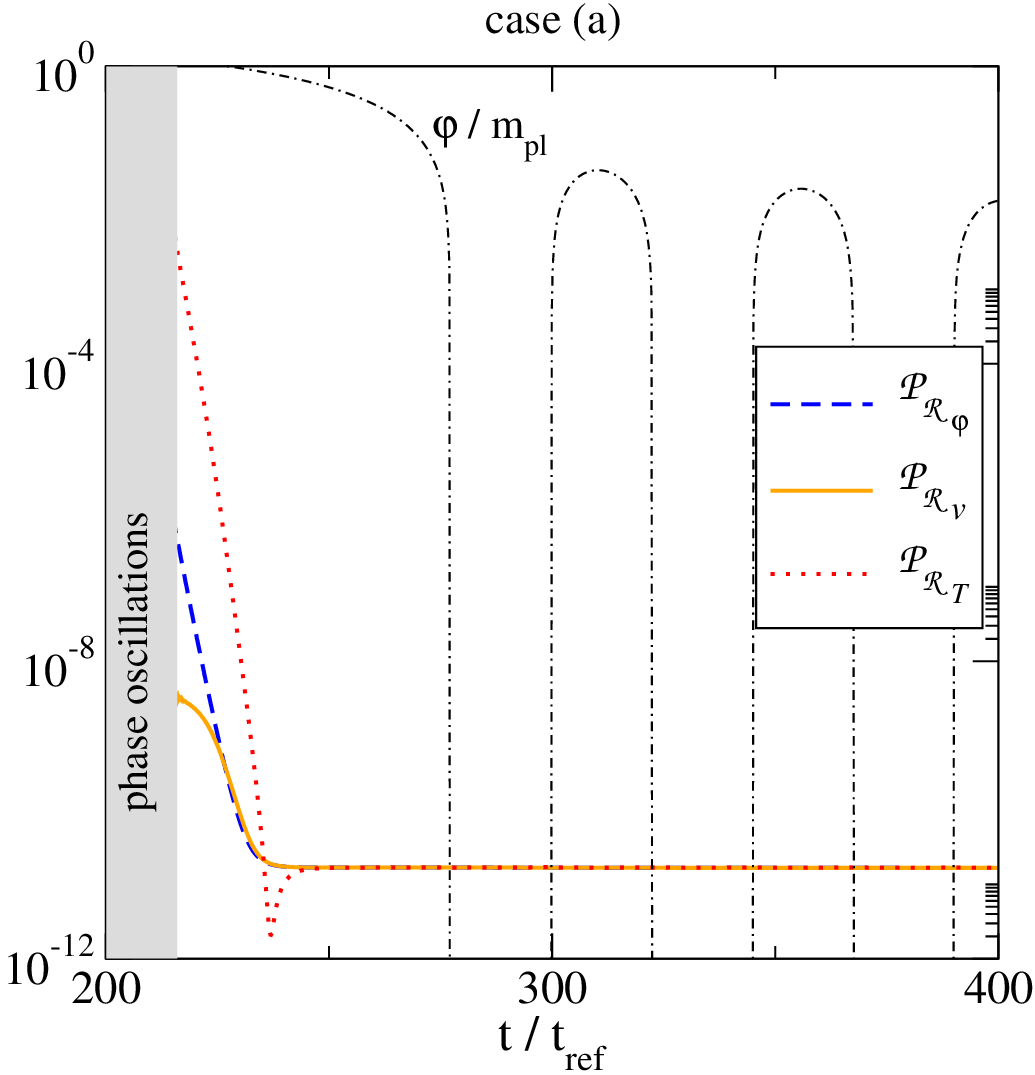}
   ~~~\epsfysize=5.0cm\epsfbox{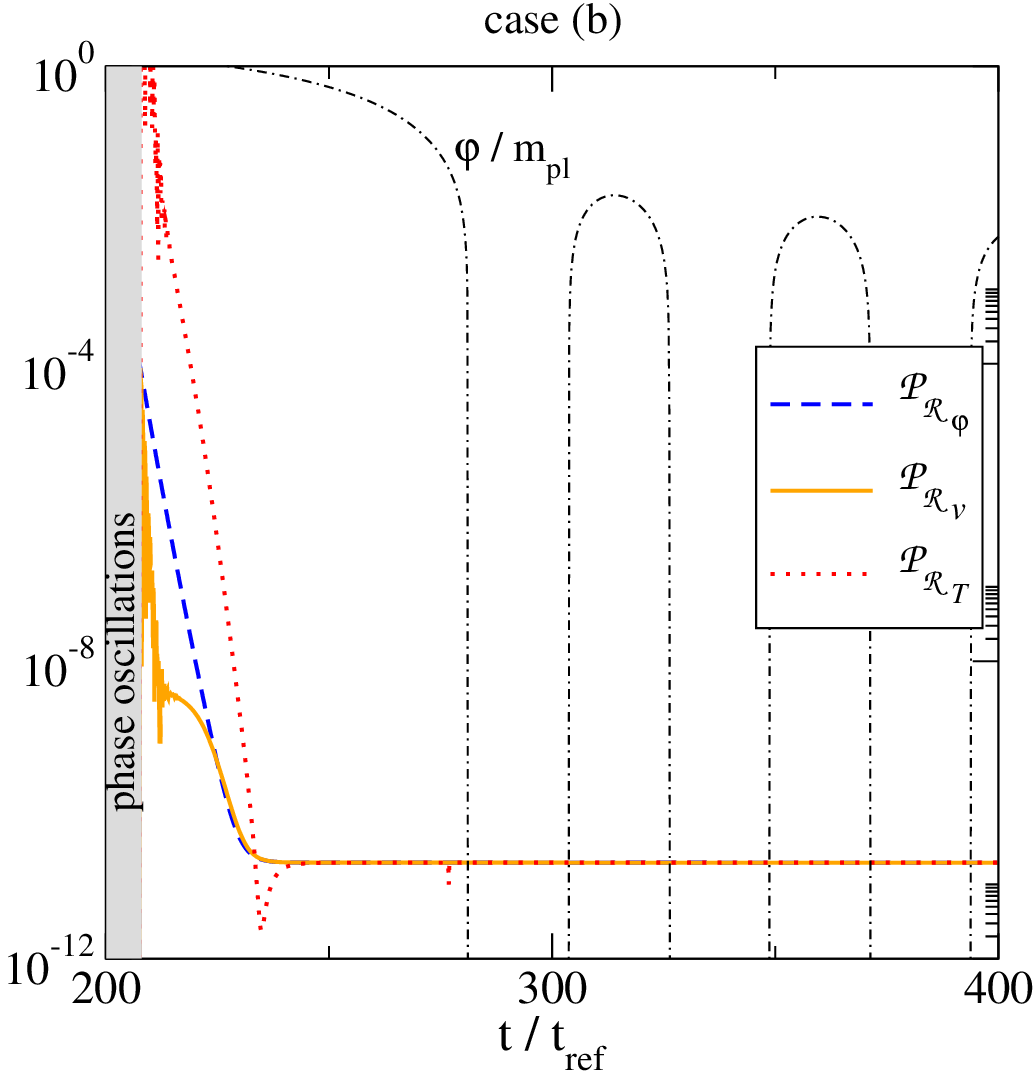}
   ~~~\epsfysize=5.0cm\epsfbox{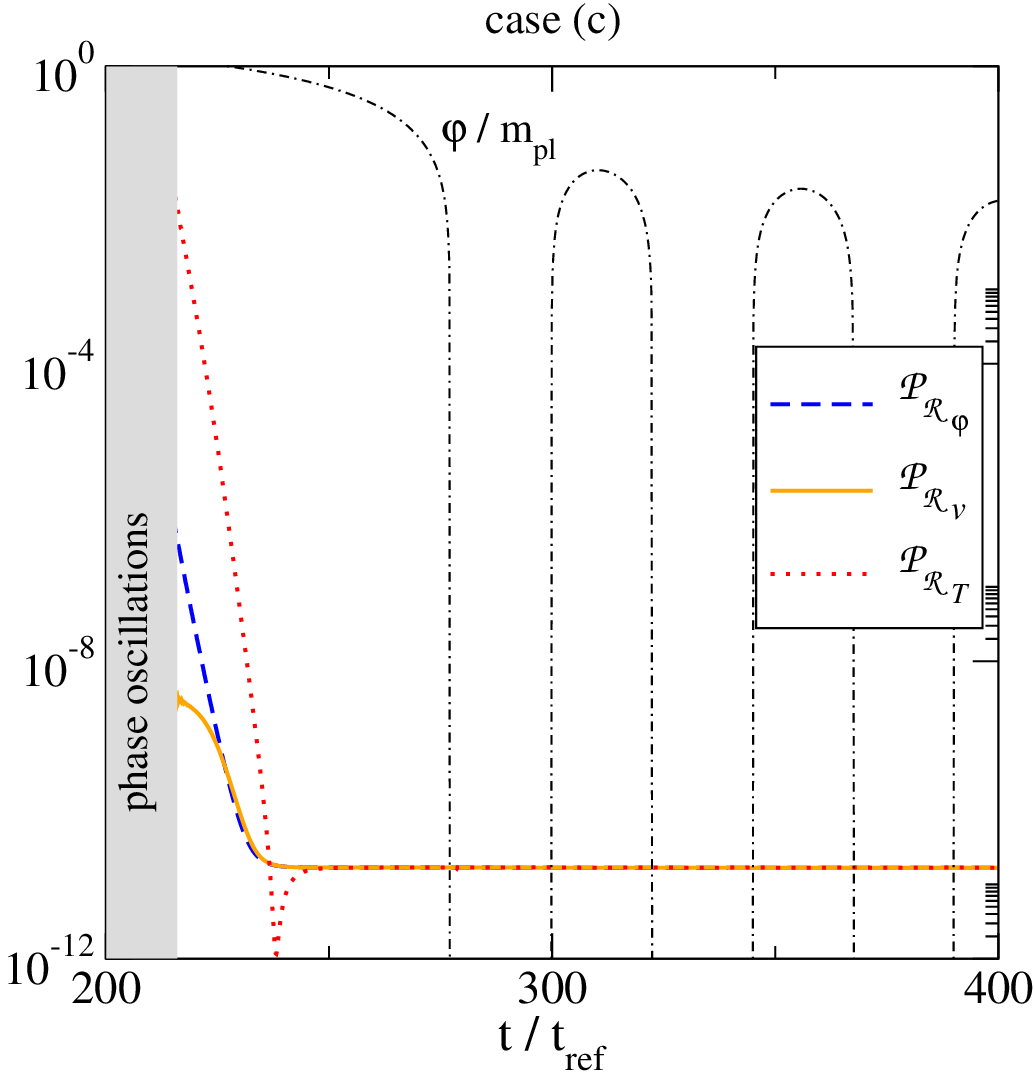}
}

\vspace*{0.3cm}

\centerline{%
   ~~~\epsfysize=5.2cm\epsfbox{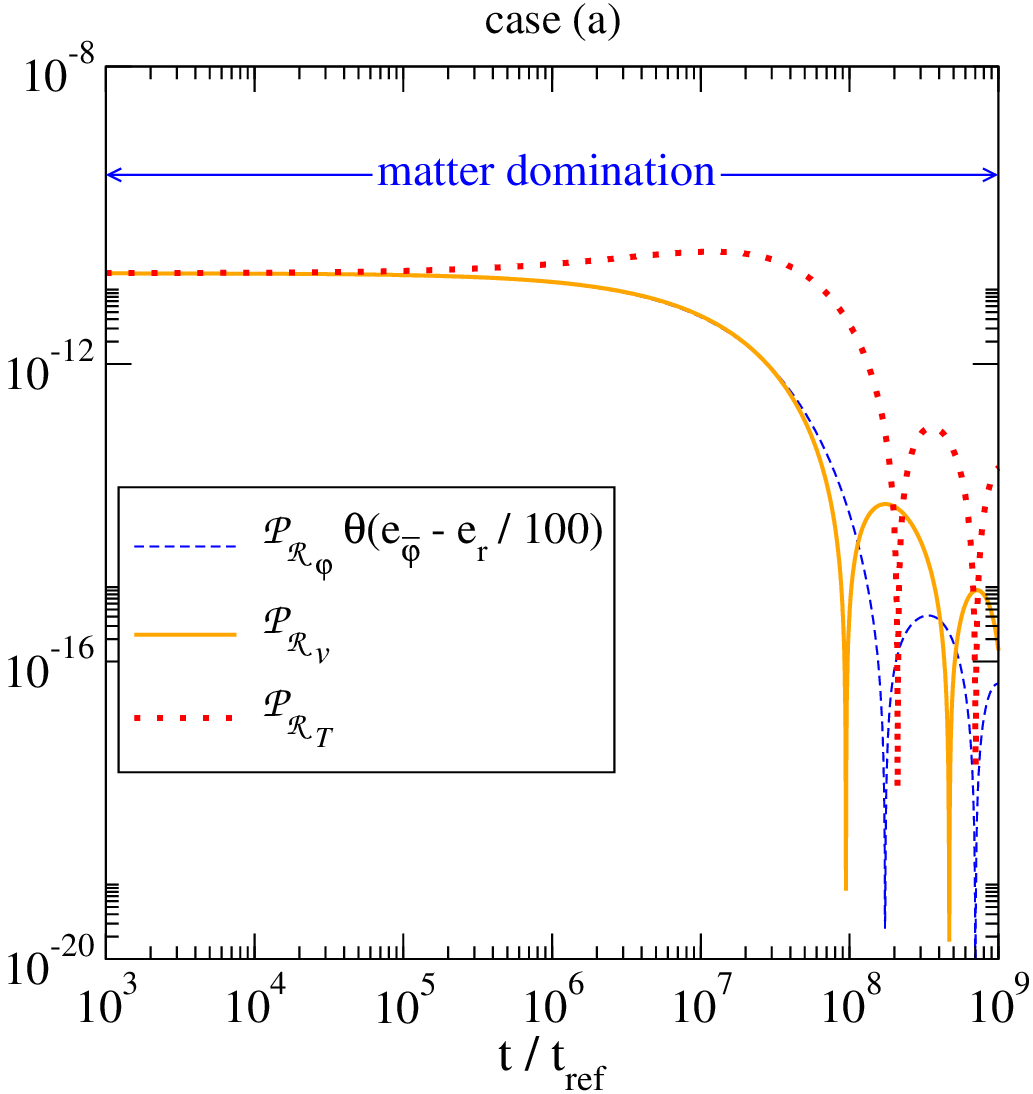}
   ~~~\epsfysize=5.2cm\epsfbox{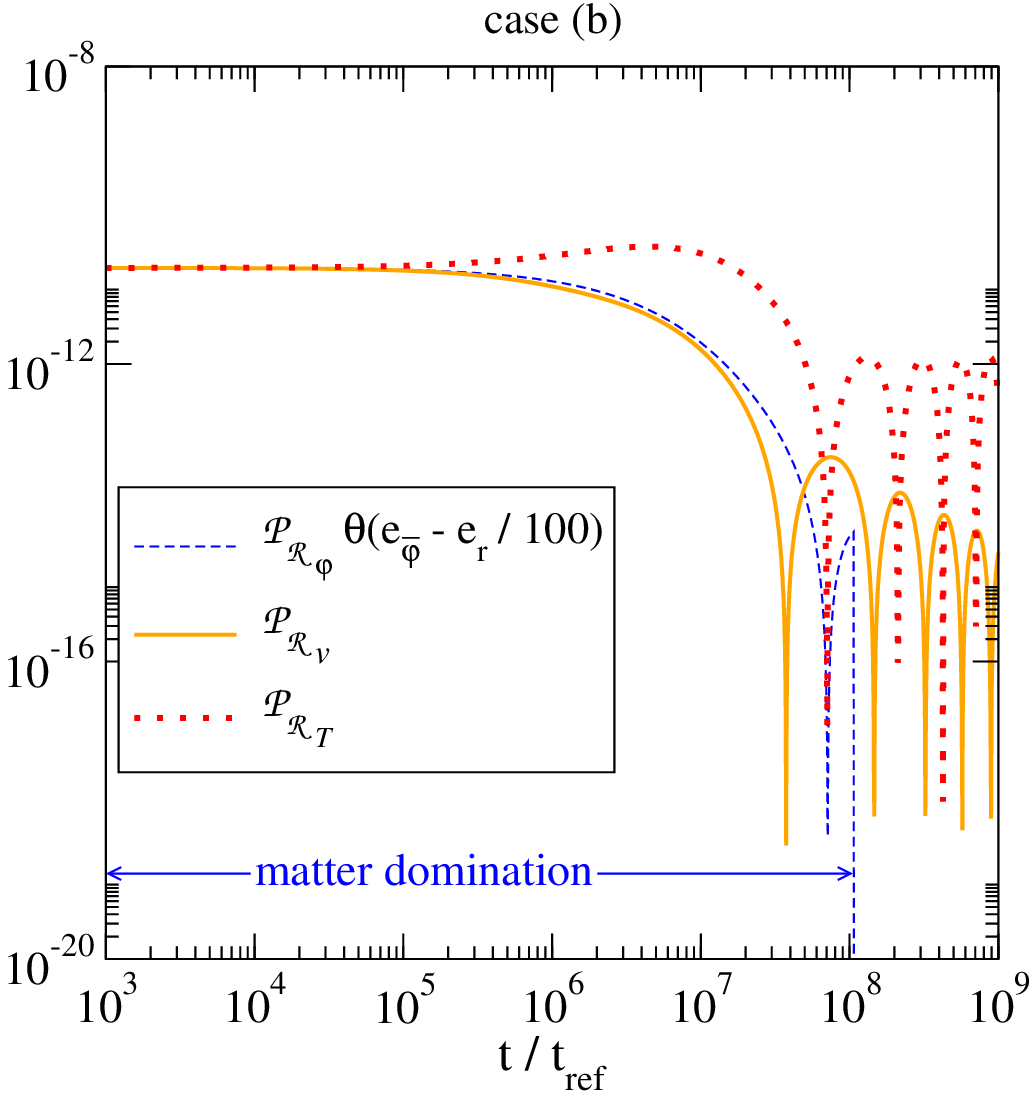}
   ~~~\epsfysize=5.2cm\epsfbox{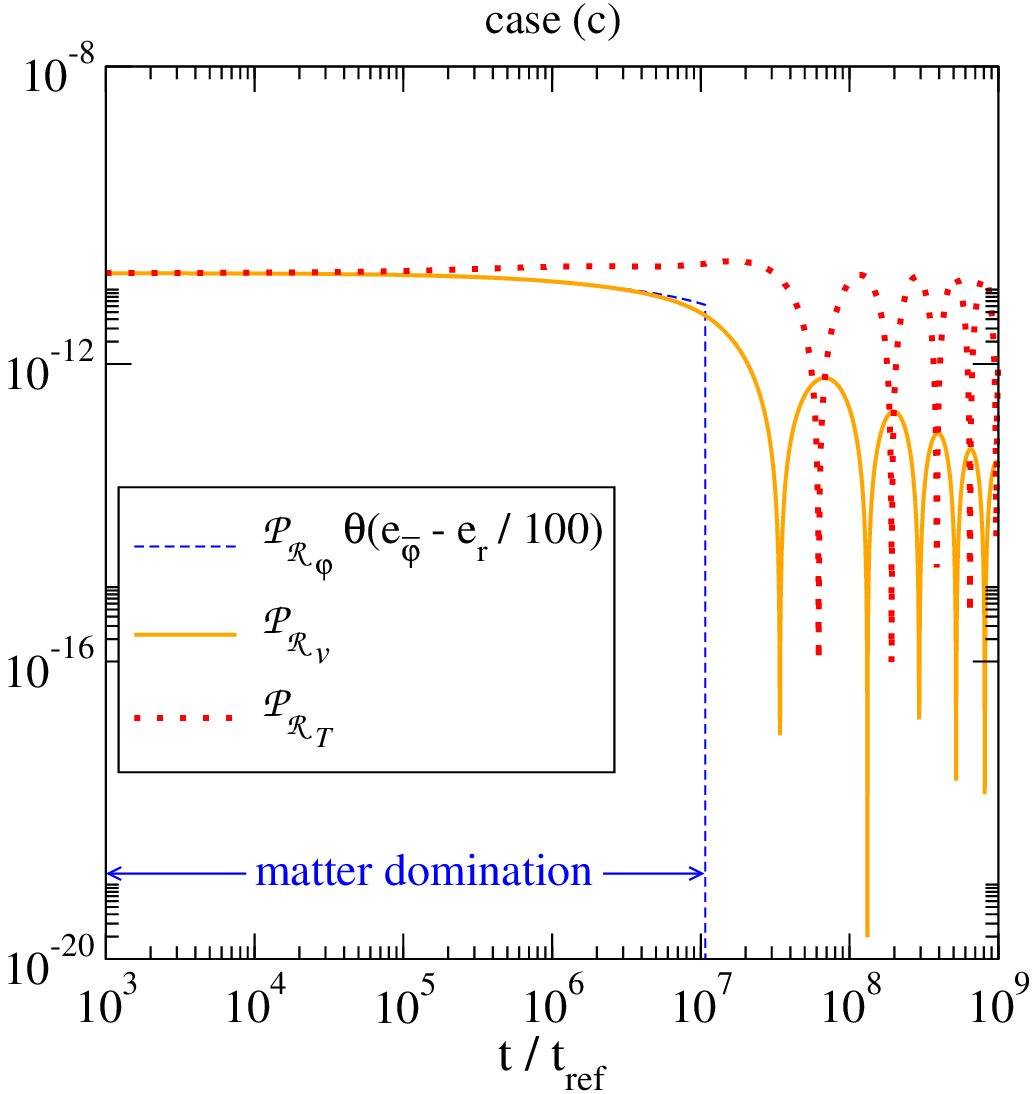}
}

\caption[a]{\small
  Examples of curvature perturbations, 
  cf.\ \eqs\nr{def_R_varphi}--\nr{def_R_T}, for the benchmarks and 
  momentum modes shown in \fig\ref{fig:cases}.
  The upper panels show the behaviour around the times when inflation ends; 
  the lower panels when background oscillations have become fast
  (\se\ref{ss:matter}) and are
  eventually overtaken by radiation (\se\ref{ss:radiation}). 
  We have set $\P^{ }_{\mathcal{R}_\varphi}\to 0$
  when we go over to the last regime.
}

\la{fig:curv}
\end{figure}

\begin{figure}[t]

\vspace*{-0.1cm}

\centerline{%
   ~~~\epsfysize=5.2cm\epsfbox{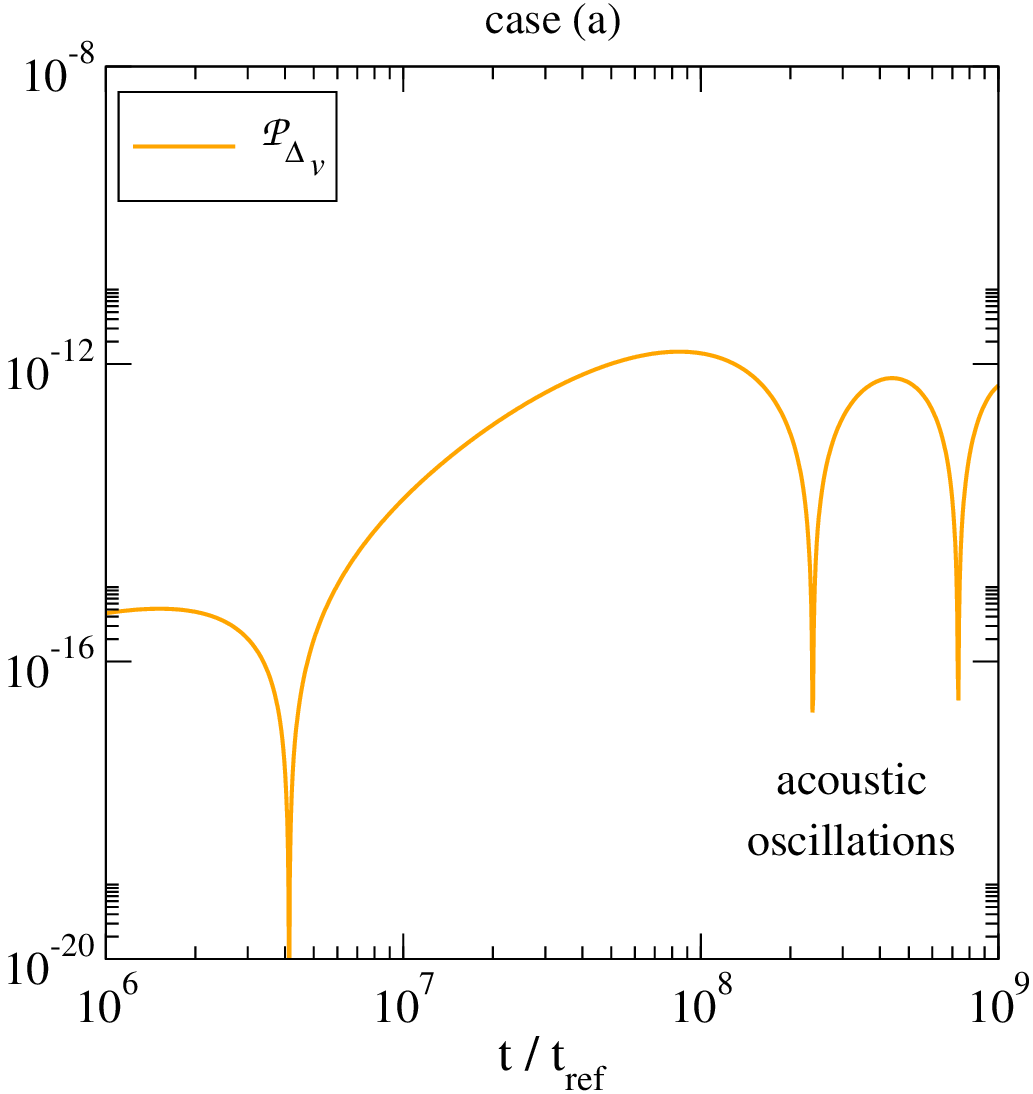}
   ~~~\epsfysize=5.2cm\epsfbox{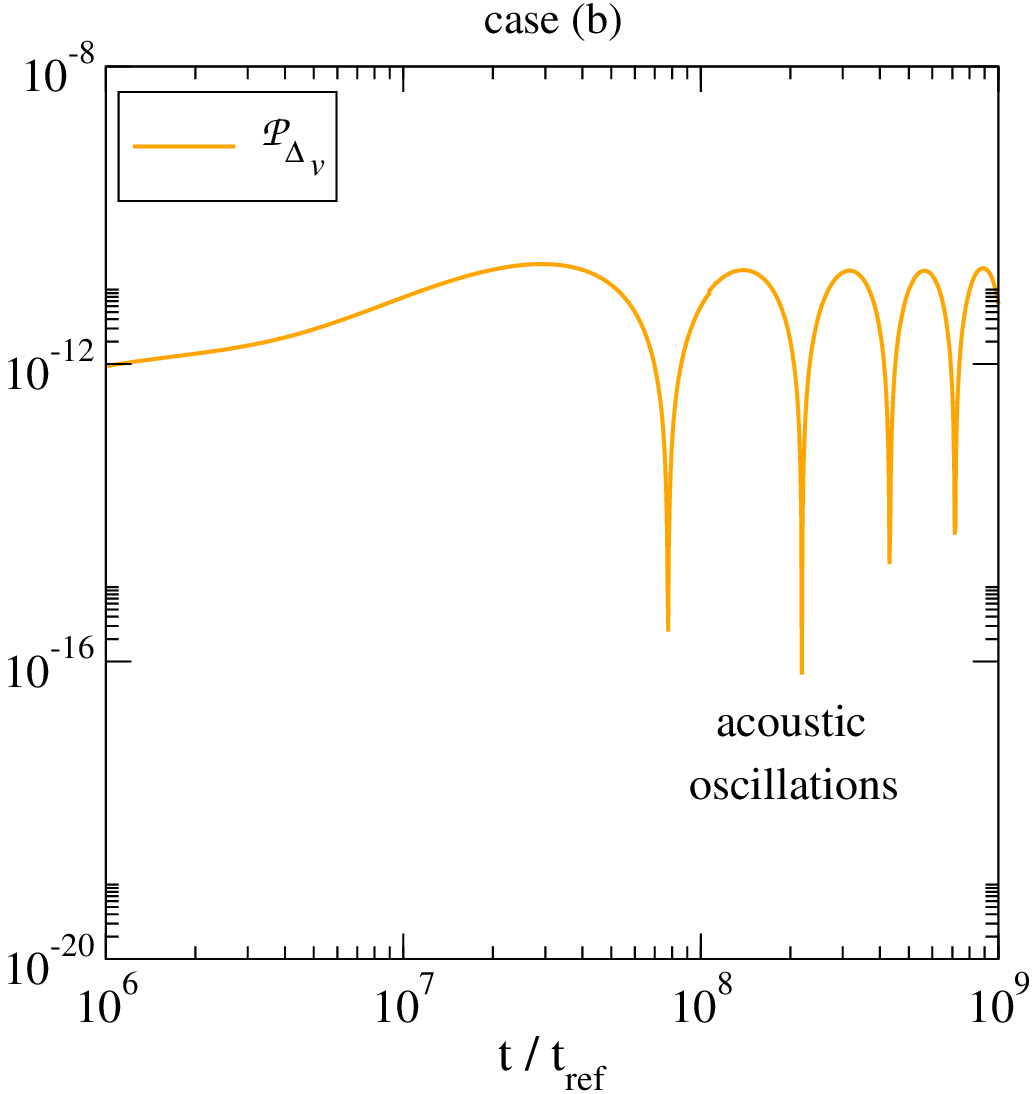}
   ~~~\epsfysize=5.2cm\epsfbox{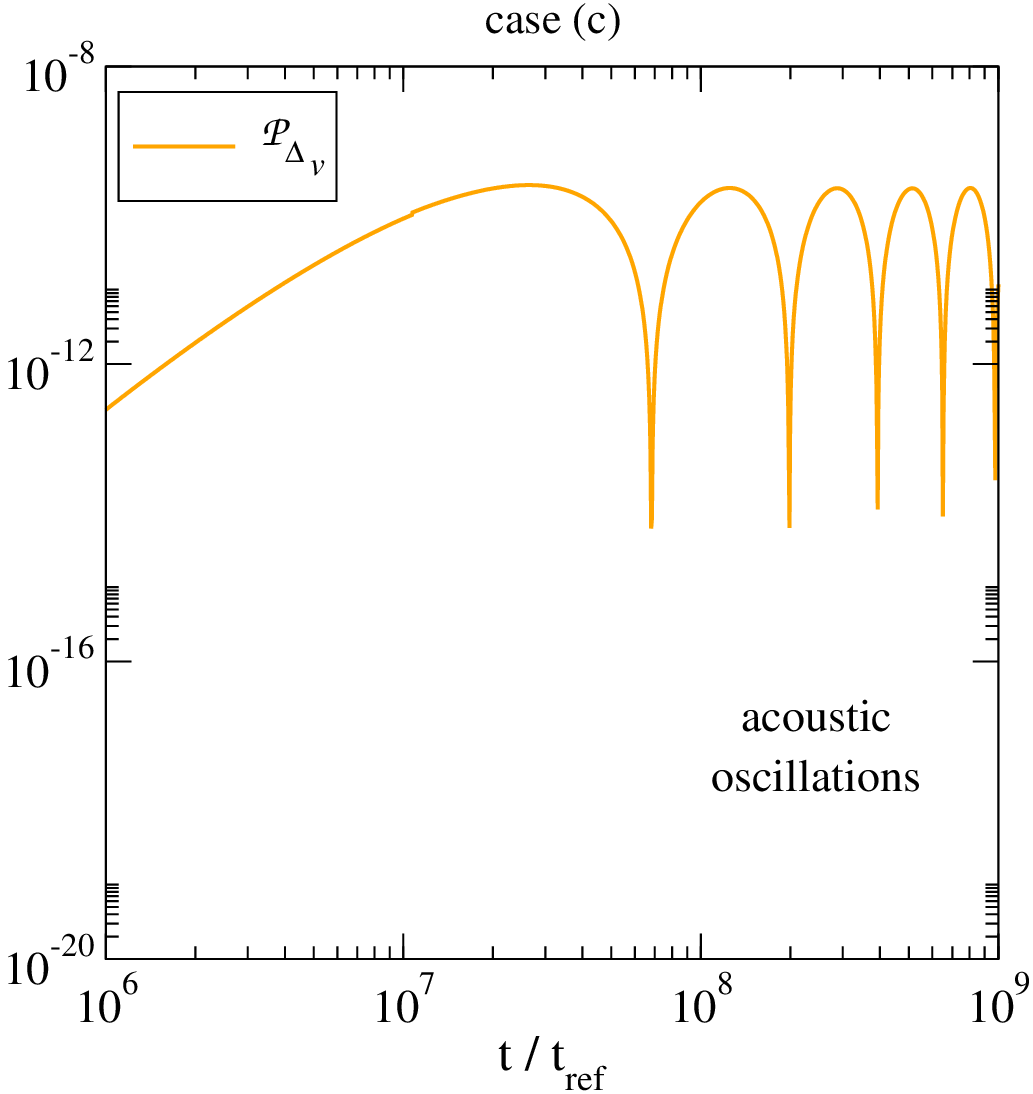}
}

\caption[a]{\small
  Examples of late-time energy density perturbations,
  from \eqs\nr{Delta_v} and \nr{Delta_v_new},  
  for the benchmarks and 
  momentum modes shown in \fig\ref{fig:cases}.
  Even if 
  the early-time curvature perturbations in \fig\ref{fig:curv}(top)
  are almost the same, there are large differences
  in the late-time
  energy density perturbations, 
  presumably reflecting different durations of the 
  matter-domination epoch, as shown in \fig\ref{fig:curv}(bottom).
}

\la{fig:ener}
\end{figure}

While most of the variables in \eqs\nr{dot_R_varphi}--\nr{dot_S_T}
had already appeared in the background solution, 
the physical momentum 
squared $k^2_{ }/a^2_{ }$ is a new quantity. 
In \eqs\nr{dot_R_varphi} and \nr{dot_S_T}, it can be seen 
next to coefficients of $\sim \rmO(H^2_{ })$.
The scale factor is given
in terms of $e$-folds as $a = a^{ }_\rmi{ref}\, e^{N}_{ }$, 
where $a^{ }_i \equiv a(t^{ }_i)$ and
$N$ is solved from $\dot{N} = H$, 
with the initial condition $N(t^{ }_\rmi{ref}) = 0$.
Going back in time, 
$k^2_{ }/a^2_{ }$ grows exponentially 
in importance, compared with 
slowly varying terms of $\rmO(H^2_{ })$. 
As explained in \se\ref{ss:initial}, we have started
the integration when $k / a^{ }_1 \approx 10^3_{ } H(t^{ }_1)$.

Numerical solutions 
are illustrated in \figs\ref{fig:curv} and \ref{fig:ener}.
The momentum mode considered is the one which reaches the 
value $k/a = 10^{-10}_{ }\,\mpl^{ }$ at $t = 10^4_{ }\,t^{ }_\rmi{ref}$.
At this time it is outside of the Hubble horizon in all cases. 
At around $t \sim 10^6_{ } t^{ }_\rmi{ref}$, 
the momentum mode comes back inside the Hubble 
horizon. Then the plasma perturbations
start to undergo acoustic oscillations. 

An important observation is 
that during the period where 
the modes are well outside of the Hubble horizon
($t/t^{ }_\rmi{ref}\approx 10^3_{ }...10^4_{ }$), all curvature
perturbations agree, 
$
 \mathcal{R}^{ }_\varphi = 
 \mathcal{R}^{ }_v = \mathcal{R}^{ }_\T 
$.
This is in accordance with the discussion below \eq\nr{dot_S_T}. 
We also note from \eq\nr{optimal_set} that~$\mathcal{R}^{ }_\T$
is obtained from the integration variables $\mathcal{R}^{ }_\varphi$
and $\E^{ }_\T$ via
$
 \mathcal{R}^{ }_\T =
 \mathcal{R}^{ }_\varphi 
 +  \E^{ }_\T / (\, 
 \bar{e}^{ }_{\der\T}\, \dot{T} \,)
$.
That~$\mathcal{R}^{ }_\T$ remains finite when the maximal 
temperature is reached, i.e.\ $\dot{T} = 0$, is a result of
the vanishing of $\E^{ }_\T$ at the same point.\footnote{%
 In practice we have regularized 
 $1/\dot{T}\to \dot{T}/(\dot{T}^2_{ } + \epsilon^2_{ })$, 
 with $\epsilon \simeq 10^{-25}_{ }\mpl^2$, however this 
 has no visible effect while avoiding spurious poles.
 }  

Finally, we note that on microscopically close inspection,
$\P^{ }_{\mathcal{R}_\varphi}$ 
in the top panels of \fig\ref{fig:curv} can be seen 
to show small spikes or jumps
when $\dot{\bar\varphi}=0$.
These are the singularities of \se\ref{ss:singularities}.  
However, the evolution is stable, i.e.\ it returns to the 
constant solution after the singularity, in accordance with 
\eq\nr{no_prblm_2}. More prominently, $\P^{ }_{\mathcal{R}_\T}$
tends to undershoot and meander before settling to a constant value, 
reflecting
an almost flat direction in the field space. However, the solution
is again stable, and finds the correct value after a while. 
We have not devoted special effort to ``curing'' these numerical issues, 
preferring rather to display their 
self-correcting nature explicitly, as an indication of the 
robustness of our basic equations. 

%
\section{Conclusions and outlook}
\la{se:concl}

The purpose of this 
paper has been to derive a coupled set of 
evolution equations for three gauge-invariant curvature
perturbations that
can be defined when an inflaton field and a thermal plasma
are simultaneously present. 
The set of variables chosen 
is given in \eq\nr{optimal_set}, and the corresponding
evolution equations in \eqs\nr{dot_R_varphi}--\nr{dot_S_T}.
Other gauge-invariant quantities, including the standard
Bardeen potentials and relative energy density
perturbations, can be obtained from a given solution,
as explained in \se\ref{ss:other}. 
Even though the same problem has been addressed
in warm inflation literature, it has been customary there 
to fix a gauge and to restrict the considerations 
to the slow-roll regime. 

We note that all 
standard ingredients of the inflationary paradigm are captured
by our equations, 
and also play a specific role in their solution: 
\bi

\item[(i)] 
In an initial slow-roll stage, 
a given high-momentum mode phase-oscillates inside the Hubble horizon.
This behaviour can be determined analytically, and can be used to set the
initial conditions for the more complicated evolution, as explained
in \se\ref{ss:initial}.

\item[(ii)] 
Subsequently, the momentum mode crosses outside the Hubble horizon. 
Then phase oscillations get overdamped, and only a constant
``zero mode'' remains present. 

\item[(iii)] 
After inflation ends, the inflaton background starts to oscillate. 
This leads to singular coefficients in the evolution equations. 
As explained in 
\se\ref{ss:singularities}, the apparent singularities can be efficiently
handled by making use of complexified variables.

\item[(iv)] 
The background oscillations become difficult
to treat when they are much faster 
than the Hubble rate. 
They can then be averaged out, as explained
in \se\ref{ss:matter}, leading to evolution equations which are 
relatively easy to solve numerically. 

\item[(v)] 
Finally, when the plasma energy density,
$e^{ }_r$, overtakes that in the oscillating inflaton,
$e^{ }_{\bar\varphi}$, the radiation-dominated era starts.
As explained in \se\ref{ss:radiation},  
$e^{ }_{\bar\varphi}$ can then
be omitted from the background equations, 
and $\mathcal{R}^{ }_\varphi$ decouples from the other perturbations.

\item[(vi)] 
When a momentum mode re-enters the Hubble horizon,
it starts undergoing acoustic oscillation, 
visible in the plasma temperature and flow velocity. 

\ei

In order to illustrate these features, we have defined a few 
benchmark scenarios in \se\ref{se:numerics}. The benchmarks are 
from a ``weak regime'', in which the friction coefficient $\Upsilon$
is smaller that the Hubble rate $H$ during inflation.
The corresponding background solutions are shown  
in \fig\ref{fig:cases}, the power spectra corresponding
to various curvature perturbations in \fig\ref{fig:curv}, 
and those for late-time energy density perturbations in \fig\ref{fig:ener}.
The plots extend until the start of 
acoustic oscillations in a radiation-dominated universe. 

Even if we have not considered concrete applications 
in this paper, we believe that our framework allows for precision 
studies of the physics of short length scales in the weak regime. 
One example is that the spectrum of scalar perturbations could 
affect the onset of a subsequent thermal phase transition
in the radiation sector~\cite{jvdv}. 
Another is the origin of scalar-induced gravitational waves, 
for which appropriate
4-point unequal-time correlation functions need to be computed
from the scalar perturbations that we have determined. 

As an outlook, we note that our equations apply also in 
a ``strong regime'', in which the friction coefficient $\Upsilon$
exceeds the Hubble rate $H$ already during the inflationary stage.
The difference is that then the noise terms, denoted by $\varrho$
in \eqs\nr{dot_R_varphi} and \nr{dot_S_T}, play an important role.
The noise terms are modified in interesting ways during stages
(iii) and (iv), as explained in \se\ref{ss:noise}. Even though
the equations remain the same as in the present study, their
numerical solution represents a challenge, as we are 
faced with stochastic differential equations. We plan to 
return to this topic in a ``part~II'' follow-up investigation, 
being then also able to compare with previous numerical
results from warm inflation literature. 
We would like to stress, however, that the
present study has been necessary, in order to establish
a foundation for that generalization.

%
\section*{Acknowledgements}

We thank Matthias Blau for nice lessons on canonical 
quantization in curved backgrounds. 
M.L.\ is grateful to Philipp Klose and Sebastian Zell, 
and S.P.\ to Marco Drewes, 
for helpful discussions.
S.P.\ is supported by the Swiss National Science Foundation 
under grant 
\href{https://data.snf.ch/grants/grant/212125}{212125}.

%
\appendix
\renewcommand{\thesection}{\Alph{section}}
\renewcommand{\thesubsection}{\Alph{section}.\arabic{subsection}}
\renewcommand{\theequation}{\Alph{section}.\arabic{equation}}


%
\section{Quantization of a scalar field in an expanding background}
\la{se:quantum}

We recall here some basics on the quantization of 
free scalar fields
(cf.,\ e.g.,\ ref.~\cite{book}). 
The computation is organized so that it is valid 
both for an expanding background and, setting $a\to \mbox{constant}$, 
in Minkowskian spacetime. 


Consider first \eq\nr{rescaled_pert_scalar_new},
in the limit that $\varrho$, $\Upsilon$, $\delta T$, and metric perturbations
are omitted:  
\be
 \delta\varphi'' + 2 \H(\tau) \delta\varphi'
 + \bigl[ a^2 \bar m^2(\tau) - \nabla^2 \bigr] \delta\varphi 
 \; 
    \overset{ \varrho,\Upsilon,\delta T \,\to\, 0 }{  
    \underset{ h^{ }_\rmiii{0},h^{ }_\rmiii{D},\nabla^2 h \,\to\, 0 }{ = } } 
 \; 
 0
 \;, \la{langevin_vac_2} 
\ee
where $ \H \; \equiv \; {a'}/{a} $, 
and the effective mass squared
$
 \bar m^2 \; \equiv \; V^{ }_{\der\varphi\varphi} |^{ }_{\varphi = \bar\varphi}
$
can depend on time. We view $\delta\varphi$
as a quantum-mechanical (Heisenberg) operator. 
Given that the equation
is linear, the quantum and classical 
time evolutions have the same form. 

It turns out to be helpful to eliminate the first-order time 
derivative from \eq\nr{langevin_vac_2}. This can be achieved be defining
$
 \delta\widehat\varphi \equiv a\, \delta\varphi
$, 
yielding (in analogy with \eq\nr{trafo_4})
\be
 \delta\widehat\varphi\hspace*{0.4mm}{}'' 
 + 
   \biggl[ a^2 \bar m^2(\tau) - \frac{a''}{a}
 - \nabla^2 \biggr] \,
   \delta\widehat\varphi 
 \; 
 = 
 \; 
 0
 \;. \la{langevin_vac_3} 
\ee

It is important to note that \eq\nr{langevin_vac_3} 
does in general {\em not}
agree with a corresponding gauge-invariant equation for 
the curvature perturbation $\widehat{\mathcal{Q}}$, 
cf.\ \eq\nr{eq_hatQ_varphi}. In particular, they contain
different effective masses squared. 
Making use of the background identities 
in \eqs\nr{bg_scalar_new} and \nr{bg_scalar_new_2}, and omitting again 
$\Upsilon$ and $T$, we find that the mass-squared
parameter appearing in \eq\nr{eq_hatQ_varphi}, 
which we denote by $\widehat{m}^2_{ }$, reads
\ba
       \widehat{m}^2_{ }(\tau) & \equiv & 
     - \frac{\H}{a\bar\varphi\hspace*{0.4mm}'}
      \biggl( \frac{a\bar\varphi\hspace*{0.4mm}'}{\H} \biggr)''_{ } 
 \nn
 & 
    \overset{ \Upsilon,T \,\to\, 0 }{  
    \underset{  }{ = } } 
 & 
    a^2 \bar m^2(\tau) - \frac{a''}{a} 
  + \underbrace{
    \frac{\H''_{ }}{\H}
  - 2\biggl( \frac{\H'_{ }}{\H} \biggr)^2_{ }
  + \frac{2 a^2_{ }V^{ }_{\der\varphi} }{\bar\varphi\hspace*{0.3mm}'}
    \, \biggl( \H - \frac{\H'_{ }}{\H} \biggr)
    }_{\rm extra~terms~(vanish~if~\it a\, = \,\rm
           const~or~\H\, = \,-1/\tau)}
 \;. \la{extras}
\ea
The extra terms are absent in 
Minkowskian and de Sitter spacetimes,
in which $\delta\widehat\varphi$ and 
$\widehat{\mathcal{Q}}$ obey the same evolution equation. 

Fortunately, 
if we consider comoving momenta $k \gg \widehat{m}$, then the precise
form of $\widehat{m}$ plays no role. This is the regime needed for 
setting the initial conditions for the variable $\widehat{\mathcal{Q}}$
(cf.\ \se\ref{ss:initial}). The discussion in the following is thus meant
to be applicable both for $\delta\widehat\varphi$ and 
$\widehat{\mathcal{Q}}$, 
and we adopt the mass squared $\widehat{m}^2_{ }$
from the context of $\widehat{\mathcal{Q}}$. 

A solution of the wave equation 
can be found with a mode expansion, 
\be
 \delta\widehat \varphi(\tau,\vec{x})
 = 
 \int \! \frac{{\rm d}^3\vec{k}}{\sqrt{(2\pi)^{3}_{ }}} \, 
 \Bigl[ \,
    w^{ }_\rmii{\vec{k}}
    \, \delta\widehat \varphi^{ }_k(\tau) 
    \, e^{ i \vec{k}\cdot\vec{x}}_{ }   
  + 
    w^{\dagger}_\rmii{\vec{k}}
    \, \delta\widehat\varphi^{\,*}_k(\tau) 
    \, e^{ - i \vec{k}\cdot\vec{x}}_{ }   
 \, \Bigr]
 \;, \la{mode_expansion}
\ee
where $k \equiv |\vec{k}|$ and the mode function satisfies 
\be
 \delta\widehat\varphi\hspace*{0.4mm}{}_k'' 
 + 
 \bigl[\, 
   k^2_{ } 
   + 
   \widehat{m}^2_{ }(\tau) 
 \,\bigr] 
 \, 
 \delta\widehat\varphi^{ }_k 
 \; 
 = 
 \; 
 0
 \;. \la{mode_functions} 
\ee
The mode function $ \delta\widehat \varphi^{ }_k $ is classical, 
and quantum mechanics lies in the operators 
$w_\vec{k}^{ }, w_\vec{k}^\dagger$. 

Now, two separate
solutions of \eq\nr{mode_functions}, 
$ \delta\widehat\varphi^{(1)}_k $ 
and 
$ \delta\widehat\varphi^{(2)}_k $,  
have the important property
that the Wronskian between them, 
\be
 \mathcal{W}[\delta\widehat\varphi^{(1)}_k,\delta\widehat\varphi^{(2)}_k]
 \; \equiv \; 
 \delta\widehat\varphi^{(1)}_k \, (\delta\widehat\varphi^{(2)}_k)'
 - 
 (\delta\widehat\varphi^{(1)}_k)' \,  \delta\widehat\varphi^{(2)}_k
 \;, \la{wronskian_def}
\ee
is independent of time: 
\be
 \mathcal{W}\hspace*{0.4mm}' 
 \; \stackrel{\rmii{\nr{wronskian_def}}}{=} \; 
 \delta\widehat\varphi^{(1)}_k \, (\delta\widehat\varphi^{(2)}_k)''
 - 
 (\delta\widehat\varphi^{(1)}_k)'' \,  \delta\widehat\varphi^{(2)}_k
 \; \stackrel{\rmii{\nr{mode_functions}}}{=} \; 0 
 \;. \la{d_wronskian}
\ee
This plays a role when we consider commutators of \eq\nr{mode_expansion}.
We assume the operators normalized as 
\be
 [\, w^{ }_{\vec{k}} , w^{\dagger}_{\vec{q}} \,]
 \; = \; 
 \delta^{(3)}_{ }(\vec{k}^{ } - \vec{q}^{ })
 \;, 
 \quad
  [\, w^{ }_{\vec{k}} , w^{ }_{\vec{q}} \,]
 \; = \; 
 [\, w^{\dagger}_{\vec{k}} , w^{\dagger}_{\vec{q}} \,]
 \; = \; 
 0\;. \la{commutators}
\ee
Making use of the fact that the mode functions only depend on 
the modulus of $\vec{k}$, and substituting 
$\vec{k}\to-\vec{k}$ in one of the terms, 
it follows that the equal-time commutator between 
a solution and its time derivative reads
\be
 \bigl[\, 
  \delta\widehat \varphi(\tau,\vec{x})
 \,,\,
  \delta\widehat \varphi\hspace*{0.4mm}{}'(\tau,\vec{y})
 \,\bigr]
 \; = \; 
 \int \! \frac{{\rm d}^3\vec{k}}{(2\pi)^{3}_{ }} \, 
 \mathcal{W}[\delta\widehat\varphi^{ }_k(\tau), 
 \delta\widehat\varphi^{\,*}_k(\tau)]
 \,
 e^{i \vec{k}\cdot(\vec{x} - \vec{y})}_{ }
 \;.
\ee
The desired canonical value,  
$
 i \, \delta^{(3)}_{ }(\vec{x} - \vec{y})
$,
is obtained if we set the Wronskian to 
\be
 \mathcal{W}[\delta\widehat\varphi^{ }_k(\tau), 
 \delta\widehat\varphi^{\,*}_k(\tau)]
 = i \quad \forall k
 \;.  \la{wronskian_value}
\ee
Given that the Wronskian is time-independent, 
this applies at all times.

As a second-order differential equation, 
the general solution of \eq\nr{mode_functions} involves 
two integration constants. Eq.~\nr{wronskian_value} only
fixes one of them. 
Concretely, if we write
$
 \delta\widehat\chi^{ }_k = 
 \alpha\, \delta\widehat\varphi^{ }_k
 + 
 \beta\, \delta\widehat\varphi^{\,*}_k
$, 
then 
$
 \mathcal{W}[\,
 \delta\widehat\chi^{ }_k, 
 \delta\widehat\chi^{\,*}_k
 \,]
 = 
 (\, |\alpha|^2_{ } - |\beta|^2_{ } \,)\,
 \mathcal{W}[\,
 \delta\widehat\varphi^{ }_k, 
 \delta\widehat\varphi^{\,*}_k
 \,]
$.
Therefore \eq\nr{wronskian_value} remains unchanged if 
$|\alpha|^2_{ } - |\beta|^2_{ } = 1$, which is known as
a Bogoliubov transformation. 

The information needed for fixing the remaining
integration constant is that 
$ \delta\widehat\varphi^{ }_k(\tau) $ should correspond
to the ``positive-energy'' or ``forward-propagating'' mode
at early times. This may become clear if we 
recall two explicit solutions 
of \eqs\nr{mode_functions} and \nr{wronskian_value}.
In Minkowskian~spacetime, where $\widehat{m} = a \bar{m}$, 
the positive-energy solution reads
\be
  a,\bar{m}\to \mbox{constant}: \quad
  \delta\widehat\varphi^{ }_k = \frac{e^{-i \hat\epsilon^{ }_k \tau}}
  {\sqrt{2\hat\epsilon^{ }_k}}
  \;, \quad 
  \hat\epsilon^{ }_k \; \equiv \; \sqrt{k^2_{ } + a^2_{} \bar{m}^2_{ }}
  \;. \la{ex_mink}
\ee
If a vacuum state is defined as 
$w^{ }_\vec{k}|0\rangle = 0 \; \forall\vec{k}$,
so that it contains no positive-energy modes, 
it is Lorentz invariant, 
given that proper Lorentz transformations do not
mix positive and negative-energy solutions. 
Simple solutions can also be found in massless
de~Sitter spacetime ($a = -1/(H\tau)$). If we pick the one 
that goes over to \eq\nr{ex_mink} for $|k\tau| \gg 1$, then it reads  
\be 
  \frac{a''}{a} \to \frac{2}{\tau^2} \;, \quad
  \bar{m}\to 0 \,: \quad
  \delta\widehat\varphi^{ }_k = \frac{e^{-i k \tau}}
  {\sqrt{2 k}}
  \biggl( 1 - \frac{i}{k\tau} \biggr)
  \;. \la{ex_ds}
\ee

In the general situation, we fix the two integration constants 
by requiring that we go over to \eq\nr{ex_mink}
at early times. 
In concrete terms, this is implemented
by imposing the initial conditions in 
\eqs\nr{initial_dot_R_varphi} and \nr{initial_R_varphi}, 
setting up a rapid phase rotation. The power spectrum
that we compute
then corresponds to an expectation value
in a distant-past vacuum, 
in the sense defined below \eq\nr{ex_mink}.
Corrections to pure phase rotations, as visible 
in \eq\nr{ex_ds}, are suppressed by 
$\sim \bigl( \frac{k}{aH} \bigr)^{-1}_{ }$, 
which in our case is chosen to be $\sim 10^{-3}_{ }$.


\small{
%

}

\end{document}